\documentclass[12pt,preprint]{aastex}

\def\etal {{\it et~al.}}
\def\msun {M_{\odot}}

\slugcomment{Submitted to Astrophysical Journal}

\shorttitle{Cygnus X-1}
\shortauthors{Vrtilek et al.}

\begin{document}


\title{
Ultraviolet observations of the X-ray photoionized wind of 
Cygnus X-1 during X-ray soft/high state\altaffilmark{1}} 

\altaffiltext{1}{Based on observations with the NASA/ESA Hubble
Space Telescope obtained at the Space Telescope Science Institute,
which is operated by the Association of Universities for Research
in Astronomy, Incorporated, under NASA contract NAS5-26555.
These observations are associated with programs GO-9646 and GO-9840.}


\author{S. D. Vrtilek$^2$, B.S. Boroson$^2$, A. Hunacek$^3$, D. Gies$^4$, C. T. Bolton$^5$}
\affil{$^1$Harvard-Smithsonian Center for Astrophysics,
60 Garden Street, Cambridge, MA 02138, $^2$Department of Physics, MIT, 
77 Massachusetts Avenue, Cambridge, MA 02139, $^3$Center for High 
Angular Resolution Astronomy, Department of Physics and
Astronomy, Georgia State University, P. O. Box 4106,  
Atlanta, GA 30302-4106, $^4$ David Dunlap Observatory, University of Toronto, P.O. Box 360,
Richmond Hill, Ontario, L4C 4Y6; Canada}
\email{svrtilek@cfa.harvard.edu}


\begin{abstract}
High-resolution ultraviolet observations of the black hole X-ray  binary 
Cygnus X-1 were obtained using the Space Telescope Imaging Spectrograph 
on the Hubble Space Telescope. Observations were taken at two epochs 
roughly one year apart;  orbital phase ranges around $\phi_{orb}$ = 0 and 
0.5 were covered at each epoch. 
We detect P Cygni line features from high (N~V, C~IV, Si~IV) 
and absorption lines from low 
(Si~II, C~II) 
ionization state material. 
We analyze the characteristics of a 
selection of P Cygni profiles and note, in particular,  a strong dependence 
on orbital phase for the high ionization material:  the profiles show 
strong, broad absorption components when the X-ray 
source is behind the companion star and noticeably weaker absorption when 
the X-ray source is between us and the companion star. 

We fit the P~Cygni profiles using the Sobolev with Exact 
Integration 
method applied to a spherically symmetric stellar wind subject to 
X-ray photoionization from the black hole. Of the wind-formed lines, the 
Si\,IV doublet provides 
the most reliable estimates of the parameters of the wind and X-ray 
illumination. The velocity $v$ increases with radius $r$ (normalized to 
the stellar radius) according to $v=v_\infty(1-r_\star/r)^\beta$, with
$\beta\approx0.75$ and $v_\infty\approx1420$~km~s$^{-1}$.
The microturbulent velocity was $\approx160$~km~s$^{-1}$.
Our fit implies a ratio of X-ray luminosity (in units of
10$^{38}$ erg s$^{-1}$) to wind mass-loss rate
(in units of 10$^{-6} \msun~yr^{-1}$) of L$_{X,38}/\dot M_{-6} 
\approx 0.33$, measured at  
$\dot M_{-6}$ = 4.8.
The lines from the lower ionization species and the 
He~II~$\lambda$1640 absorption are consistent with formation in the 
photosphere of the
normal companion.

Our models determine parameters that may be used to estimate the accretion
rate onto the black hole and independently predict the X-ray luminosity.
Our predicted L$_x$ matches that determined by contemporaneous RXTE ASM
remarkably well, but is a factor of 3 lower than the rate according to
Bondi-Hoyle-Littleton spherical wind accretion.  We suggest that some of the
energy of accretion may go into powering a jet.

\end{abstract}




\section{Introduction}

Cygnus X-1, the only Galactic X-ray binary with a high mass companion 
where existing
observations require a black hole for the compact object, was first 
discovered in 1962
(Cowley 1992, and references therein).  
In addition to the well-established high mass function found
with optical observations, the
X-ray data of Cyg X-1
display transitions from a high flux state in the 2-10 keV band
(where a strong soft component
dominates) to a low flux state (where the soft 
component largely disappears) 
that has been interpreted as characteristic of black hole systems. 
It also displays highly broadened Fe K$\alpha$ emission (Miller~\etal~2002) 
that is consistent
with models for X-ray reflection in Galactic black holes and AGNs. The
broad-line shape of Fe K$\alpha$ may be caused by  
Doppler shifts and the gravitational field of the black hole.

The Cyg X-1 system consists of a supergiant star and a compact object.
The mass of the compact object is in the range 
7-20 $\msun$ (Shaposhnikov \& Titarchuk 2007; Ziolkowski 2005)
the mass of the 
visible star is in the range  
18-40$\msun$ (Ziolkowski 2005; Tarasov~\etal~2003; Brocksopp~\etal~1999). 
The binary orbital period 
is 5.6 days (Bolton 1972).
Miller~\etal~(2005) using Chandra/HETG observations
find that the X-ray spectrum of Cygnus X-1 at phase 0.76 is 
dominated by absorption lines, in strong contrast to
spectra of other HMXBs such as Vela X-1 and Cen X-3.
Schulz~\etal~(2002) report marginal evidence for ionized
Fe transitions with P-Cygni
type profiles at orbital phase 0.93 whereas Marshall~\etal~(2001) find
no evidence for such line profiles at phase 0.84.
Miller~\etal~(2005)  
suggest that, while the spectra of Cen X-3 can be modelled
by a spherically-symmetric wind (Wojdowski~\etal~2003), the X-ray
absorption spectrum of
Cyg X-1 requires dense material preferentially along the line of
sight;  considered together, the Chandra spectra provide 
evidence in X-rays for a focused wind in Cygnus X-1.

Our Space Telescope Imaging Spectrometer (STIS) observations of Cygnus X-1 
were obtained when Cyg X-1 was in its soft/high X-ray state and
show line profiles that change 
significantly between orbital phases 0.0 and 0.5. 
We interpret these changes in terms of models that
include the effects of X-ray photoionization
on the stellar wind of the normal companion or of a focused wind.
We test our model predictions with contemporaneous X-ray observations.
The observations and analysis are described in Section 2,
our models of the line profiles are presented in Section 3, and
our interpretation and conclusions are discussed in Section 4.
\\

\section{Observations and Analysis}

\begin{figure}
\includegraphics[width=1.0\textwidth]{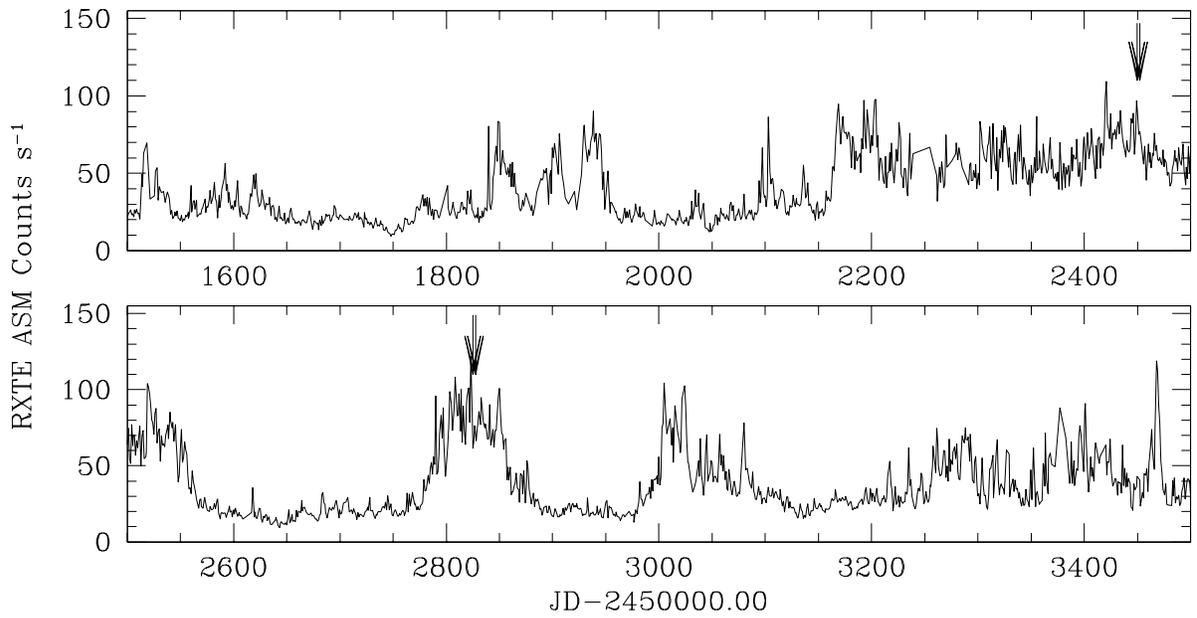}
\caption{One day averages of the flux observed from Cyg X-1
with the All-Sky Monitor on board the Rossi X-Ray Timing Explorer.
Arrows indicate the times of the HST/STIS observations.} 
\label{rxtehst}
\end{figure}

Cyg X-1 was observed with the STIS on the Hubble Space Telescope (HST)
when the X-ray source was behind the normal star and half an
orbit later on two separate epochs roughly one year apart.
Figure \ref{rxtehst} shows the times of our observations in comparison with
the 1 day--averaged light curves obtained from the all-sky monitor
on the Rossi X-Ray Timing Explorer (RXTE; Levine~\etal~1996).  
The RXTE light
curves show  that the HST observations were taken when the binary was at 
relatively high X-ray flux which is associated with the X-ray soft/high 
state. 

The STIS instrument design and in-orbit performance have been described 
by Woodgate et al. (1998) and Kimble et al. (1998). The E140M grating 
provided a resolving power of 6 km s$^{-1}$ in the wavelength range 
1150--1740$\AA$. 
The data were processed through the standard HST/STIS
pipeline and further reduced using the STSDAS routines available through IRAF.

Table 1 lists the HST observation identifier,
along with 
the start dates, exposure length, 
and orbital phase at the start of each observation. Phases were calculated using
an ephemeris that places phase zero at JD2441874.707$\pm$0.009, 
with an orbital period of
5.599829$\pm$0.000016 (Brocksopp~\etal, 1999).
Phase zero corresponds to
supergiant inferior conjunction.

\begin{figure}
\includegraphics[width=1.0\textwidth]{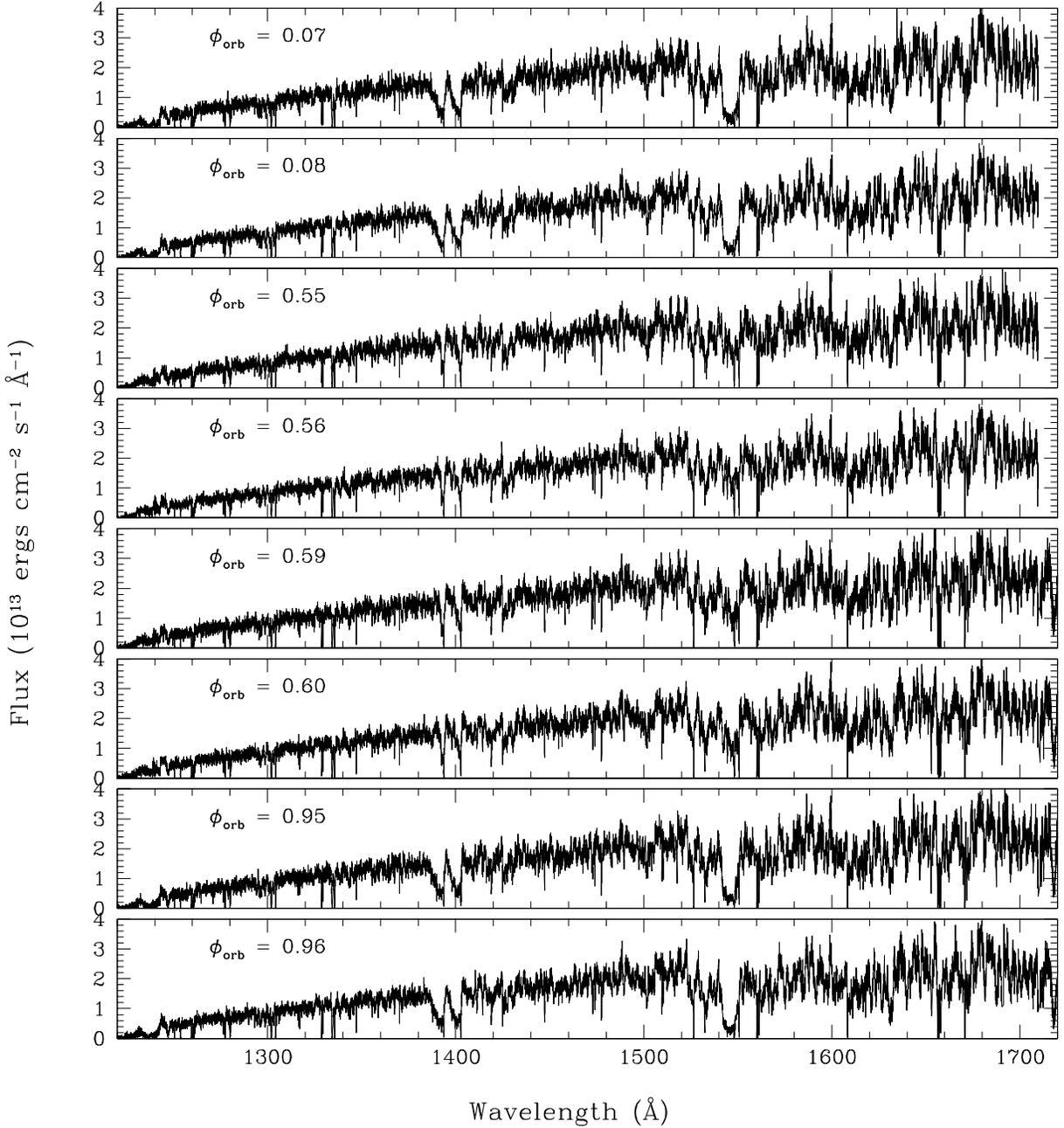}
\caption{Unsmoothed, raw, HST/STIS data of Cygnus X-1 stacked with
increasing orbital phase, regardless of epoch.}
\label{cygx1hstall}
\end{figure}

Figure \ref{cygx1hstall} shows the eight datasets in increasing orbital
phase order regardless of epoch.  These data are not smoothed and not
corrected for reddening and no instrumental quality control has been
applied (i.e., removal of hot pixels, etc).  
It is clear that the changes in high ionization 
material line profiles are orbital
phase dependent, with stronger absorption near orbital phase 0.0.
The close-ups of the N~V, Si~IV, and C~IV regions in Figure 3 demonstrate
that this orbital phase dependence persists for observations taken a
year apart.

Figure \ref{cygx1lines} shows a spectrum taken at orbital phase 0.96,
with the stronger spectral lines labeled. Table 2 lists the
features identified using the program SpecView 
\footnote{$http://www.stsci.edu/resources/software\_hardware/specview$}
and tables supplied by NIST  
\footnote{$http://www.physics.nist.gov/PhysRefData/Handbook/index.html$}. 
We note that some of the lines
are saturated and several are blended with other features.  
Figures \ref{cygx1nv}-\ref{cygx1heii} show profiles of both 
high (N~V, C~IV, Si~IV) and 
low (Si~II, C~II) ionization state material in each of 
two orbital phases at two epochs.  
The low ionization material and He~II are likely to be from the stellar
surface rather than the wind.
\\

\section{Line Profile Models}
We have implemented a model that uses the Sobolev method with Exact
Integration (SEI; Lamers~\etal~1987) to predict the P Cygni line profiles
from a wind ionized by an embedded X-ray source. The details of the wind
ionization are identical to those given by Boroson~\etal~(1999; equations
11-16). The SEI method extends the escape probability method of McCray 
\etal~(1984) by including integration along the line of sight. This allows
scattering from a range of points around the ``resonant point", as will
result when there is small-scale turbulence (microturbulence).  The SEI
method also allows a more exact treatment of interaction between P~Cygni
line doublet components.

We take note of a complementary analysis of the same STIS observations by
Gies~\etal~(2007). Gies~\etal~also use the SEI method to
analyze the wind, but assume that the ions that form the P~Cygni lines
are {\it only} present in the region of the wind in which the primary
blocks X-ray ionization from the black hole.

Following McCray \etal~(1984), we compute the local ionization fraction 
by combining ambient photoionization (from shocks in the wind, etc.)
that would be present in an isolated O~star with the ionization rate 
computed using the XSTAR code, assuming a local ionization parameter 
$\xi=L_x/n r_x^2$ (with X-ray luminosity $L_x$, number density $n$, and 
distance from the black hole $r_x$). We use the X-ray spectrum shape
modeled by Wilms~\etal~(2006) for observations taken near our HST
observations.  
Where the O~star shadows the wind 
from X-ray photoionization by the black hole, we assume only the ambient 
photoionization rate, which gives optical depths in the line 
parameterized by $\alpha_1$ and $\alpha_2$, or $\tau_{wind}$, often used 
for winds of isolated OB stars (Lamers~\etal~1987). The XSTAR ionization 
code is described in Bautista \&\ Kallman (2001).

We fit the model, which includes adjustable free parameters, to the 
spectra at phases near both superior and inferior conjunction of the 
black hole. Thus, the fits are sensitive to both the blue-shifted 
absorption trough and the red-shifted emission peak. Although the lines 
are saturated, the optical depths of the lines in the absence of X-ray 
photoionization still affect the resultant fits.

We attribute the sharp features (velocity widths of order 30 km~s$^{-1}$) 
to interstellar absorption lines, the narrow absorption lines (widths of 
94 km~s$^{-1}$) originate on the stellar photosphere, and the broadest 
features come from the wind (terminal velocity from our model of 1420 
km~s$^{-1}$).  In addition, when we model the N\,V lines, we incorporate 
narrow interstellar Mg\,II lines at $1239.925,1240.397$\AA. The SEI 
calculation of the line profiles near the rest velocity can require 
integration to start extremely close to the stellar surface in order to 
agree with the results of comoving frame methods (Lamers~\etal~1987). 
Hence we expect the P~Cygni line fits to be least reliable near the rest 
wavelengths of the lines.

We assumed fixed values for parameters describing the O~star and orbit as
given in Table 3.
Our results are not sensitive to the abundance values
used, except for the determination of $\dot{M}$.
The parameters describing the wind and X-ray
illumination are allowed to vary until a $\chi^2$ minimum is found through
the downhill simplex (``amoeba") method of Nelder \& Mead (1965).
The errors we quote are the rms variation in the final simplex, in which 
the $\chi^2_\nu$ varies by less than 1\%. There may be other local 
$\chi^2$ minima and some of the parameters may interact. For example, 
different values of $\tau_{\rm wind}$, $\alpha_1$, $\alpha_2$ may produce 
similar functions of optical depth versus radius in the wind.

The Si\,IV doublet is generally the most reliable indicator of wind
behavior in stars of this spectral type (Lamers~\etal~1999), as the N\,V
and C\,IV lines are saturated. We show that the lines can be fit using a
wind acceleration law that is standard for OB stars: 
$v=v_\infty(1-1/r)^{\beta}$, with $r$ the radius in the wind normalized to
the stellar radius, and $v_\infty$ and $\beta$ free parameters of our
model. (We note that our code used $v=0.01 v_\infty + 0.99*v_\infty(1-1/r)^{\beta}$ in
order to avoid singularities).  The best-fit value of $\beta$ is $\approx0.75$, which is standard
for OB star winds. The best-fit value of $v_\infty\approx1420$~km~s$^{-1}$
is lower than the 2300~km~s$^{-1}$ found by Davis \& Hartmann (1983). 
It is possible that the higher terminal velocity found
by Davis \& Hartmann was due to line blending on the blue side
of the feature that was hard to discern in the weak IUE spectrum.
We also note that Davis \& Hartmann used IUE data of mostly C~IV to determine
their terminal velocity whereas we use STIS data of Si~IV.  Also we allow
for microturbulence.
For the C\,IV lines we fix $\beta$ to that found for the Si\,IV fits, and 
then for N\,V, which has a lower signal to noise ratio, we fix also 
$v_\infty$. The best-fit parameters for the fits shown in 
Figures \ref{nvmod}-\ref{civmod} are given in Table 4.

\section{Accretion Rate
and X-ray Luminosity}

\subsection{Mass Accretion Rate} 
The fits to the changing P~Cygni lines by a model of X-ray ionization
within a spherically symmetric wind determine parameters that may be
used to estimate the accretion rate onto the black hole. For accretion
purely through capture of stellar wind material, Bondi \&\ Hoyle (1944)
show that the rate at which mass is captured from a stellar wind by a
compact object ($\dot M_{\rm capture}$) is given by
\begin{equation}
\dot M_{\rm capture}=\frac{4 \pi G^2 M_{\rm bh}^2 \rho}{V_{\rm rel}^3}
\end{equation}
where $G$ is the gravitational constant, $M_{\rm bh}$ is the mass of the
compact
object, $\rho$ is the density of the undisturbed gas flow near the compact
object, and $V_{\rm rel}=(v_{\rm wind}^2+v_{\rm orbit}^2)^{1/2}$ is the
velocity of the wind relative to the compact object, which has velocity
$v_{\rm orbit}=2\pi R/P$ for orbital radius $R$ and period $P$ in a wind
we take to have $v_{\rm
wind}=v_\infty(1-R_*/R)^\beta$. From mass conservation we have
\begin{equation}
\rho=\frac{\dot{M}}{4 \pi R^2 v_{\rm wind}}
\end{equation}

We can relate $\dot M_{\rm capture}$ to the X-ray luminosity if we assume
accretion releases energy with some efficiency $e\approx0.1$, so that
$L_x=e \dot M_{\rm capture} c^2$. For disk accretion, we expect $e=0.057$
for a nonrotating black hole and $e=0.42$ for a maximally rotating black
hole (Shapiro \&\ Teukolsky 1983, p. 429).

Thus our best-fit values for $\dot{M}$, $\beta$, and $v_{\infty}$ 
provide an independent prediction of $L_x$.  This prediction is based on
the idealization that the accretion proceeds entirely through the
gravitational capture of stellar wind material. The prediction also
depends on other uncertain parameters ($e$, $M_{\rm bh}$).  Our
estimate, $L_x\approx8\times10^{37}$~erg~s$^{-1}$ depends sensitively on
the velocity of the stellar wind near the black hole. As $v_\infty$
varies from $1000$ to $2000$~km~s$^{-1}$, $L_x$ goes from
3$\times10^{37}$ to $3\times10^{38}$~erg~s$^{-1}$.

The model dependence on the properties of the
X-ray ionized region of the wind also leads to a determination
of 
$L_x/\dot{M}$ and $\dot{M}$, which together give
$L_x=1.6\times10^{37}$~erg~s$^{-1}$ and
$\dot M = 4.8\times10^{-6}\msun~yr^{-1}$ in the fit to the Si\,IV
lines. The C\,IV and N\,V lines, which we expect to be less reliable,
give $L_x=1.8\times10^{38}$ and $L_x=8.0\times10^{37}$~erg~s$^{-1}$,
respectively.

We can compare these models for the X-ray luminosity with the observed
contemporaneous RXTE ASM count rate, $\approx80$~counts~s$^{-1}$
(Figure~\ref{rxtehst}). Schulz et al. (2002) observed Cygnus X-1 with
{\it Chandra} during a period in which the RXTE ASM showed flares. An
upper estimate for the ASM count rate during their observation,
$\approx50$~cts~s$^{-1}$, together with their measure of the 0.5-10 keV
X-ray luminosity as 1.6$\times10^{37}$~erg~s$^{-1}$ (for a distance of
2.5 kpc), would imply an X-ray luminosity of $\ge
2.6\times10^{37}$~erg~s$^{-1}$ during the STIS observations.

\subsection{X-ray Luminosity}

One further uncertainty in our model of the wind is that we assume a
constant X-ray luminosity whereas the X-rays from Cygnus X-1 often flare
and display a complex power spectrum. The saturated P~Cygni lines may
respond nonlinearly to flares, although light travel times may diminish
the response. In Figures~\ref{changlum1} and \ref{changlum2}, we
show how the P~Cygni lines may change in our model in response to a
change in X-ray luminosity. If the change were linear, the graph of the
model with best-fit value of $L_{\rm x}$ would be identical to the
average of the other two graphs.
Whereas we see a clear asymmetry.

In Figures \ref{conplot1}, \ref{conplot2}, and
\ref{conplot3}, we show contours of constant values of 
Log$_{10}(a_{\rm Si\,IV})$, where $a_{\rm Si\,IV}$ is the fraction of Si that is
in the form Si\,{\sc IV}. The lines extending radially from the center
of the O~star show the lines of sight at $\phi=0.55$ (extending to the
bottom) and $\phi=0.96$ (extending to the top), adjusted for orbital
inclination $i$ by the use of an effective phase $\phi^\prime$ such that
$\cos 2 \pi \phi^{\prime}=\cos 2 \pi \phi \sin i$.

Figures \ref{conplot1} through \ref{conplot3} show that the
black hole is very effective at removing the Si\,{\sc IV} ion from the
wind. However, it is difficult to see intuitively the optical depth in
the wind, given the ion fraction.
Therefore, we show in Figures~\ref{contau1} through
\ref{contau3} the radial optical depths of Si\,{\sc IV} in the wind,
given the same three X-ray luminosities, $L_x=(1/3, 1, 5/3)\times
1.6\times 10^{37}$ erg~s$^{-1}$, with other parameters fixed to the
best-fit values.  From these plots it should be apparent that the global
ionization in the wind is sensitive to the X-ray luminosity. This is
particularly true at the low end of the X-ray luminosity range. In that
case, a large region outside of the X-ray shadow remains at an optical depth
that causes noticable signatures in the line profiles, in spite of
X-rays from the black hole.

In these figures, we use our parameterization of the background optical
depth in terms of $\alpha_1$, $\alpha_2$, and $\tau_{\rm wind}$ to
determine the optical depth in the shadow region. The resulting contours
of constant optical depth are ellipsoidal, compacted towards the compact
object. 

\section{Discussion and Conclusions}

The Space Telescope Imaging Spectrograph on Hubble provides the highest
resolution ultraviolet spectra taken of Cyg X-1 to date.  Observations
were taken at two epochs roughly a year apart:  at each epoch
orbital phases when the compact object is behind the stellar 
companion and when
the compact object is in front of the companion star were covered.   
We find P~Cygni profiles from high ionization (N\,V, C\,IV, Si\,IV) gas.
For both epochs the P~Cygni profiles
show significantly less absorption at phases when the compact object
is in the line of sight. RXTE observations indicate that the X-ray flux
of the system was at a similar level at each epoch.  The observed
changes can be attributed entirely to orbital effects.  We interpret 
this to mean that X-rays from the compact object photoionize the 
wind from the massive companion
resulting in reduced absorption by the wind material.  

P Cygni profiles of selected species are consistent with the
Hatchett-McCray effect, in which X-rays from the compact object
photoionize the stellar wind from the companion star, thereby reducing
absorption. This effect also appears in UV observations of LMC X-4 and SMC
X-1 (Boroson et al. 1999; Vrtilek et al. 1997; Treves et al. 1980). 
SEI models can fit the observed P Cygni profiles and
provide measurements of the stellar wind parameters. The Si\,IV fits are
the most reliable and we use them to determine L$_x/\dot{M} =4.8 \pm0.3
\times 10^{42}$ ergs s$^{-1}$ M$_{\odot}^{-1}$ yr, where L$_{X}$ is
the X-ray luminosity and $\dot{M}$ is the mass-loss rate of the star. The
results from the C\,IV and N\,V lines are less reliable because they are 
saturated and the
C\,IV fit does not match the data well.  For these fits we fixed the
terminal velocity $\nu_{\infty}$ and the microturbulent velocity to those
given by our fits to Si\,IV. 

The best fit values for the optical depth in the ambient wind are high
($\ge 10$).  Once saturated, the OB star wind lines hardly change with
large optical depth, but when much of the wind is ionized by the black
hole, the ion fraction in the remaining regions can have a significant
effect on the line profile.

The reduced absorption when the compact object is in the line of sight
is inconsistent with focusing of
the wind toward the compact object, as has been suggested by several
authors (e.g, Sowers~\etal~1998; Tarasov~\etal~2003; Miller~\etal~2005),
as then we would expect more absorbers in the line-of-sight and hence
increased P Cygni absorption. 
Also, IUE observations taken at 8 orbital phases show a continuous
variation in the P Cygni profiles with maximum absorption at phase 0.0
and minimum at 0.5 (Treves~\etal~1980; van Loon~\etal~2001). 
Further high spectral resolution ultraviolet observations of Cyg X-1 will
be necessary to 
study the behavior of the P Cygni lines during different 
X-ray states. 

In an analysis of 2 years of RXTE/ASM data
Wen~\etal~(1999) found
the 5.6 day orbital period of Cyg X-1 during the X-ray low/hard state,
but no evidence
of the orbital period during the high/soft state.
Wen \etal~suggest that absorption of X-rays by a stellar wind from the
companion star can reproduce the observed X-ray orbital modulations in
the hard state: The lack of modulation in the soft state could be due
either to a reduction of the wind density during the soft state or to
partial covering of a central hard X-ray emitting region by an
accretion stream.
Gies~\etal~(2003) used the results of a four year spectroscopic monitoring
program of the H$\alpha$ emission strength of
HDE226868, the normal companion to Cyg X-1, to argue that the
low/hard X-ray state occurs when there is a strong, fast wind and accretion
rate is low, while in the high/soft state a weaker, highly ionized
wind attains only a moderate velocity and the accretion rate increases.
The interpretations of both Wen~\etal~(1999) and Gies~\etal~(2003)
are inconsistent with the fact that the {\it total} X-ray luminosity
from 1.3-200 keV remains constant during both the X-ray soft and hard
states (Wen~\etal~1999): the designation of X-ray high or X-ray low 
during these states is
only applicable
for the
1.5-12 keV ASM
band. 
Since the 1.3-200 keV X-ray luminosity is unchanged from the hard to 
soft state,  
fluctuations in the narrow ASM band cannot be due to reduction in 
accretion,
or obscuration of the X-ray source; rather it is a physical change that
causes the dominant emission mechnism to switch from thermal to power-law.

We note that the orbital modulation observed by Wen~\etal~in the hard
state ($\pm 1.6$ ASM cts/sec around the average) is less than the errors on
the counts during the soft state ($\pm 3$ ASM cts/sec);
we suggest that the quality of the RXTE ASM is not sufficient
to detect this low level orbital modulation during the soft state.
Ultraviolet observations clearly show
orbital modulation during both X-ray hard and soft states (Gies~\etal~2007).
Our wind models explain both the X-ray and ultraviolet flux during the
soft/high state.  We need high spectral resolution ultraviolet observations
during the hard/low state to determine if there is a change in wind
density between states.

Our determination of the mass-accretion rate can be considered as a 
positive check on the Hatchett-McCray
models, as the Si\,IV line gave $L_x=1.6\times10^{37}$~erg~s$^{-1}$, and
the model is subject to systematic uncertainties. However, this value of
$L_x$ is a factor of 3 lower than the best estimate of the accretion
rate according to Bondi-Hoyle-Littleton spherical wind accretion. 
We suggest that some of the energy of accretion may go into powering the jet.

Our test of the dependence of our results on X-ray luminosity 
confirms the utility of our models and
demonstrates that the wind outside the shadow zone is still sensitive to
X-ray illumination, an arrangement which allows us to fit $L_{\rm x}$ as
a free parameter in our model.

In future observations of the time-variability of the wind lines, the
light travel-time effects may be used to advantage, as the wind may act
as a ``low-pass filter" to the X-ray observations, with the filter
cutoff indicating the size of the ionized region (Kallman, McCray, \&\
Voit 1987).

\acknowledgments
AH would like to acknowledge support from the
REU program NSF grant 9731923 awarded to SAO. DG would like to
acknowledge support provided by NASA
through a grant (GO-9840) from the Space
Telescope Science Institute, which is
operated by the Association of Universities for Research in Astronomy,
Incorporated, under NASA contract NAS5-26555. The X-ray results were
provided by the ASM/RXTE teams at MIT and at the RXTE SOF and GOF at
NASA's GSFC. The contour graphs of ion fraction benefitted from the
programming of Corey Casto, who helped increase the resolution of the
contours.



\clearpage

\begin{table*}
\begin{center}
{\bf Table 1: Log of Observations}
\vskip 0.1in
\begin{tabular}{lcccc}
\hline
\hline
Obs. ID:&&&Exposure&Orbital\\
HST Archive ID&Obs. Start Time&MJD&(s)& Phase$^*$\\
\hline
\hline
Dataset 1: O8HD1010&2002-06-24 18:10:02&52449.757&2676.00&0.55\\
Dataset 2: O8HD1020&2002-06-24 19:36:41&52449.817&2172.00&0.56\\
Dataset 3: O8HD2010&2002-06-27 16:39:48&52452.694&2172.00&0.07\\
Dataset 4: O8HD2020&2002-06-27 18:06:34&52452.754&2676.00&0.08\\
Dataset 5: O8NX1010&2003-07-05 04:27:23&52825.185&2147.00&0.59\\
Dataset 6: O8NX1020&2003-07-05 05:51:58&52825.244&2675.00&0.60\\
Dataset 7: O8NX2010&2003-07-07 04:27:47&52827.185&2147.00&0.95\\
Dataset 8: O8NX2020&2003-07-07 05:53:29&52827.245&2675.00&0.96\\
\hline
\end{tabular}
\end{center}
\noindent
$^*$Phase computed at center of observation.
\end{table*}

\begin{table}
\begin{center}
\centerline{\bf Table 2: Line identifications for the spectrum of Figure
4.}
\vskip 0.1in
\begin{tabular}{lccc}
\hline
\hline
&Laboratory&&\\
Line ID&Wavelength ($\AA$)&Transition&Comment\\
\hline
\hline
Ly$\alpha$&1215.670&1-2           $_{1/2-*}$ &saturated\\
N~V&1238.821&$^2$S-$^2$P$^o$ $_{1/2-3/2}$ &saturated\\
N~V&1242.804&$^2$S-$^2$P$^o$ $_{1/2-1/2}$&saturated\\
Si~II&1260.422&$^2$P$^o$-$^2$D           $_{1/2-3/2}$&\\
O~I&1302.168&$^3$P-$^3$S$^o$           $_{2-1}$  &weak,blended with Si~I
I\\
O~I&1304.858&$^3$P-$^3$S$^o$           $_{1-1}$   &weak,blended with Si~
II\\
Si~II&1304.370&$^2$P$^o$-$^2$S          $_{1/2-1/2}$&weak,blended with O
~I\\
Si~II&1309.276&$^2$P$^o$-$^2$S          $_{3/2-1/2}$&weak,blended with O
~I\\
C~II&1334.532&$^2$P$^o$-$^2$D           $_{1/2-3/2}$ &weak\\
C~II&1335.663&$^2$P$^o$-$^2$D          $_{3/2-3/2}$ &weak\\
Si~IV&1393.755&$^2$S-$^2$P$^o$          $_{1/2-3/2}$&\\
Si~IV&1402.770&$^2$S-$^2$P$^o$         $_{1/2-1/2}$&\\
Si~II&1526.707&$^2$P$^o$-$^2$S          $_{1/2-1/2}$&\\
Si~II&1533.431&$^2$P$^o$-$^2$S          $_{3/2-1/2}$&\\
C~IV&1548.203&$^2$S-$^2$P$^o$          $_{1/2-3/2}$&\\
C~IV&1550.777&$^2$S-$^2$P$^o$          $_{1/2-1/2}$&\\
He~II&1640.420&2-3             *-* &weak\\
Al~II&1670.787&$^1$S-$^1$P$^o$           $_{0-1}$&\\
\hline
\end{tabular}
\end{center}
\end{table}

\begin{table}
\begin{center}
{\bf Table 3: Model Parameters}
\vskip 0.1in
\begin{tabular}{lccc}
\hline
\hline
Symbol & Adopted value & Meaning & Comments\\
\hline
\hline
$i$ & 40$^\circ$ & Orbital inclination
& $<20-67^{\circ}$ (Gies~\etal~2003, Tarasov~\etal~2003;\\
&&& Wen~\etal~1999; Herrero~\etal~1995;\\
&&&Friend \&\ Cassinelli 1986;
Davis \& Hartmann 1983)\\
$R_{\rm O}$ & $1.5\times10^{12}$~cm &
Radius of O star & $1.23\times10^{12}$~cm (Davis \&\ Hartmann 1983)\\
&&&  $1.2-1.6\times10^{12}$~cm (Gies \& Bolton 1986)\\
$R_{\rm orbit}$ & 2.0 $R_{\rm O}$ & Semimajor axis & 2.4
$R_{\rm O}$
(Davis \&\ Hartmann 1983)\\
$V_{sin i}$ & 94~km~s$^{-1}$ &
(Projected  & $94.3\pm5$
(Ninkov et al. 1987),  \\
&&rotation velocity)& 96 (Gies \&\ Bolton 1986) \\
$v_{\rm sys}$ & 0 &
Systemic velocity & 0$^{+6}_{-9}$ km~s$^{-1}$ (Ninkov
et al. 1987)\\
$a_{\rm C}$ & 3.6$\times10^{-4}$ & Carbon
abundance & solar from Verner~\etal~1994\\
$a_{\rm N}$ & 1.1$\times10^{-4}$ & Nitrogen
abundance & solar from Verner~\etal~1994\\
$a_{\rm Si}$ & 3.5$\times10^{-5}$ & Silicon abundance & solar from Verner
~\etal~1994\\
\hline
\end{tabular}
\end{center}
\end{table}

\begin{table}
\begin{center}
{\bf Table 4:  Best-fit values for free parameters for Cygnus X-1 P~Cygni line
 fits}
\vskip 0.1in
\begin{tabular}{lccc}
\hline
\hline
Symbol & Fitted value (Si\,IV, C\,IV, N\,V) & Definition$^*$&  Comments\\
\hline
\hline
$\beta$ & 0.748$\pm0.003$ & Wind acceleration & Fit to
Si\,IV,\\
&&&
then fixed\\
$\tau_*$  & (0.436$\pm0.007$, 0.88$\pm0.03$,
0.63$\pm0.05$) & Optical
depth of stellar &
\\
&&absorption line&\\
$\tau_{\rm wind}$ & (12.7$\pm0.2$, 25.2$\pm0.1$, 7.8$\pm1.0$) & Wind total
depth
&
\\
$v_{\infty}$ & (1420$\pm10$, 1330$\pm30$, 1430) km s$^{-1}$ & Wind
terminal
velocity & 2300
km~s$^{-1}$
(Davis  \\
&&& \&\ Hartmann 1983), \\
&&& Frozen for N\,V\\
$v_{\rm turb}$ & (160$\pm10$, 292$\pm1$, 200) km s$^{-1}$ & Wind
microturbulence & Frozen for
N\,V\\
$\alpha_1$ & (-0.31$\pm0.03$, 0.32$\pm0.01$, 0.66$\pm0.10$) & Wind opacity 
exponent & \\
$\alpha_2$ & (0.89$\pm0.06$, 0.140$\pm0.004$, 0.80$\pm0.01$) & Wind opacity 
exponent& \\
$L_{\rm x,38}/\dot{M}_{-6}$ & (0.033$\pm0.001$, 0.38$\pm0.01$,
0.12$\pm0.01$) & Ratio, X-ray
luminosity &\\
&&(10$^{38}$ erg s$^{-1}$)
to wind &\\
&& mass loss ($10^{-6}$
M$_\odot$ yr$^{-1}$)
& \\
$\dot{M}_{-6}$ & (4.8$\pm0.3$, 4.8$\pm0.1$, 6.7$\pm0.2$) & Wind
mass loss rate &\\
&&(10$^{-6}$ M$_{\odot}$ yr$^{-1}$) &\\
$\chi^2_\nu$ & (3.0, 5.8, 1.3) & Goodness of fit & \\
$\nu$ & (933, 596, 585) & Degrees of freedom & \\
\hline
\end{tabular}
\end{center}
$^*$From Lamers, Cerruti-Sola, \& Perinotto (1987).
\end{table}

\clearpage

\clearpage
\begin{figure}
\includegraphics[width=1.0\textwidth]{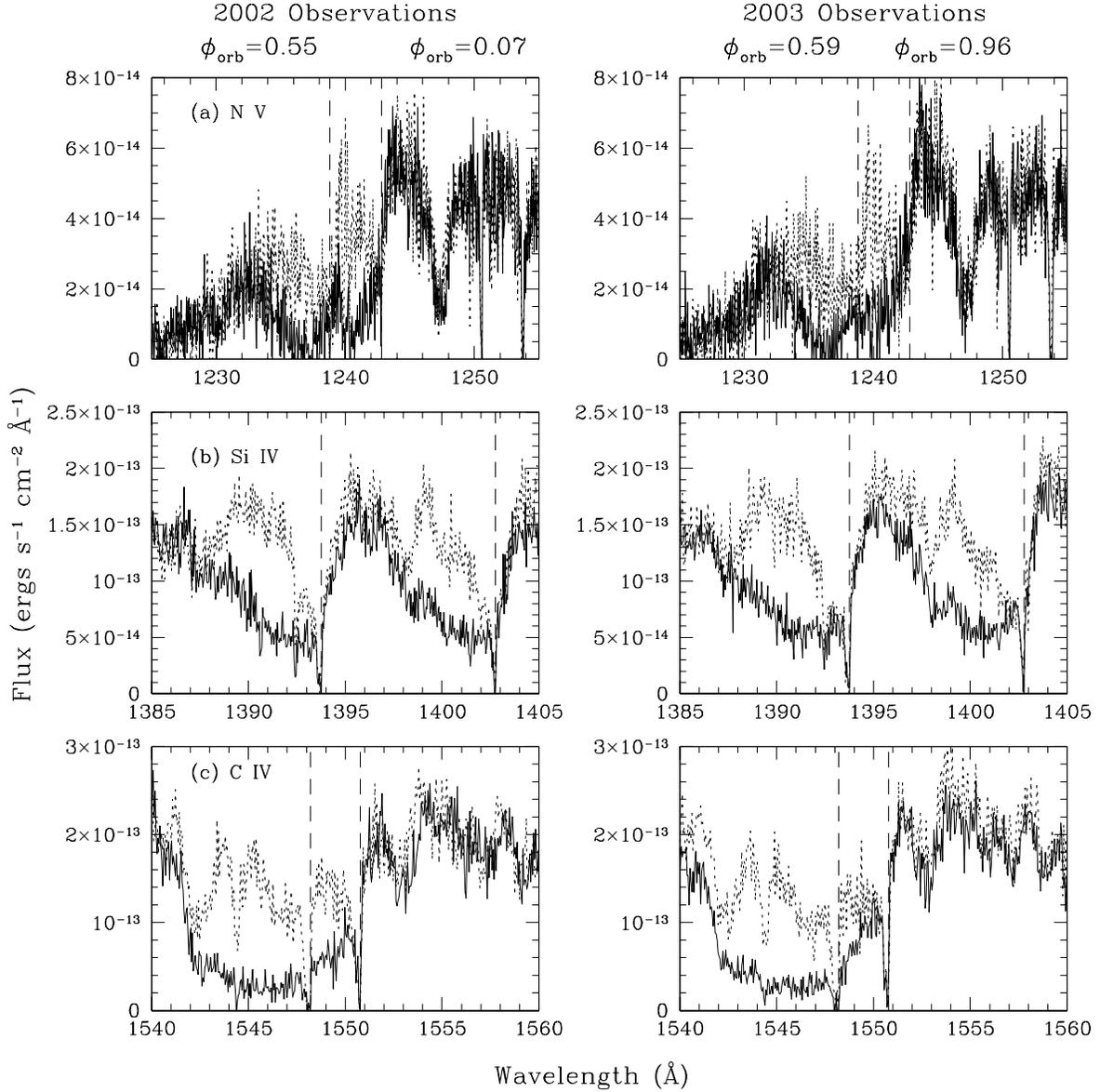}
\caption{
STIS observations of Cyg X-1
taken roughly a year apart
showing enhanced absorption at orbital phases when the X-ray source is
behind the companion star (phase 0.0; solid lines) and illustrating the stability of the absorption
change as a function of orbital phase.  Narrow features are interstellar.  Black
dashed lines indicate the laboratory wavelengths of transitions identified in Table 2.
}
\label{cygx1pcygni}
\end{figure}

\begin{figure}
\includegraphics[width=1.0\textwidth]{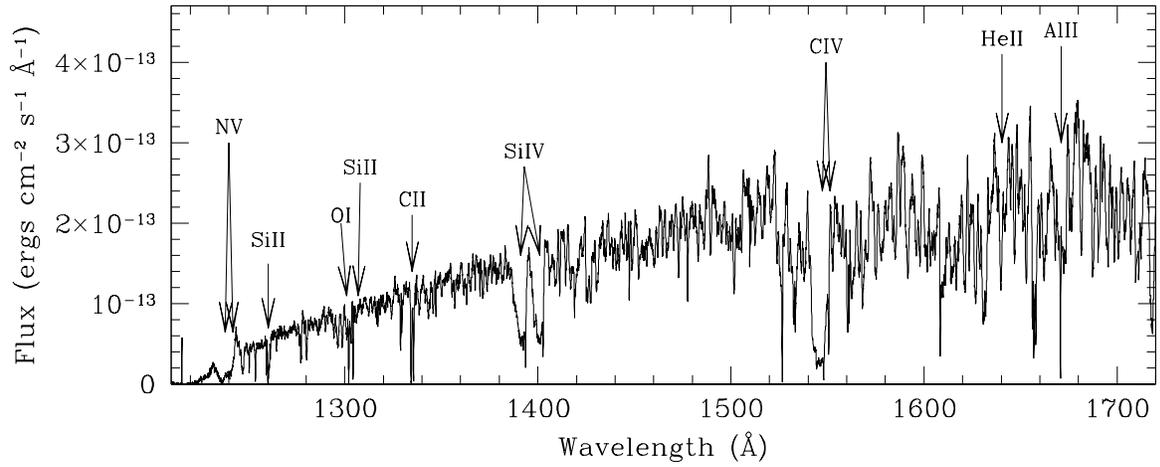}
\caption{A Cyg X-1 STIS spectrum at phase 0.96 with major
  spectral features indicated.  The data have been smoothed with a
7-point boxcar function to better show the features of interest. 
Sharp features are interstellar absorption lines.}
\label{cygx1lines}
\end{figure}

\begin{figure}
\includegraphics[width=1.0\textwidth]{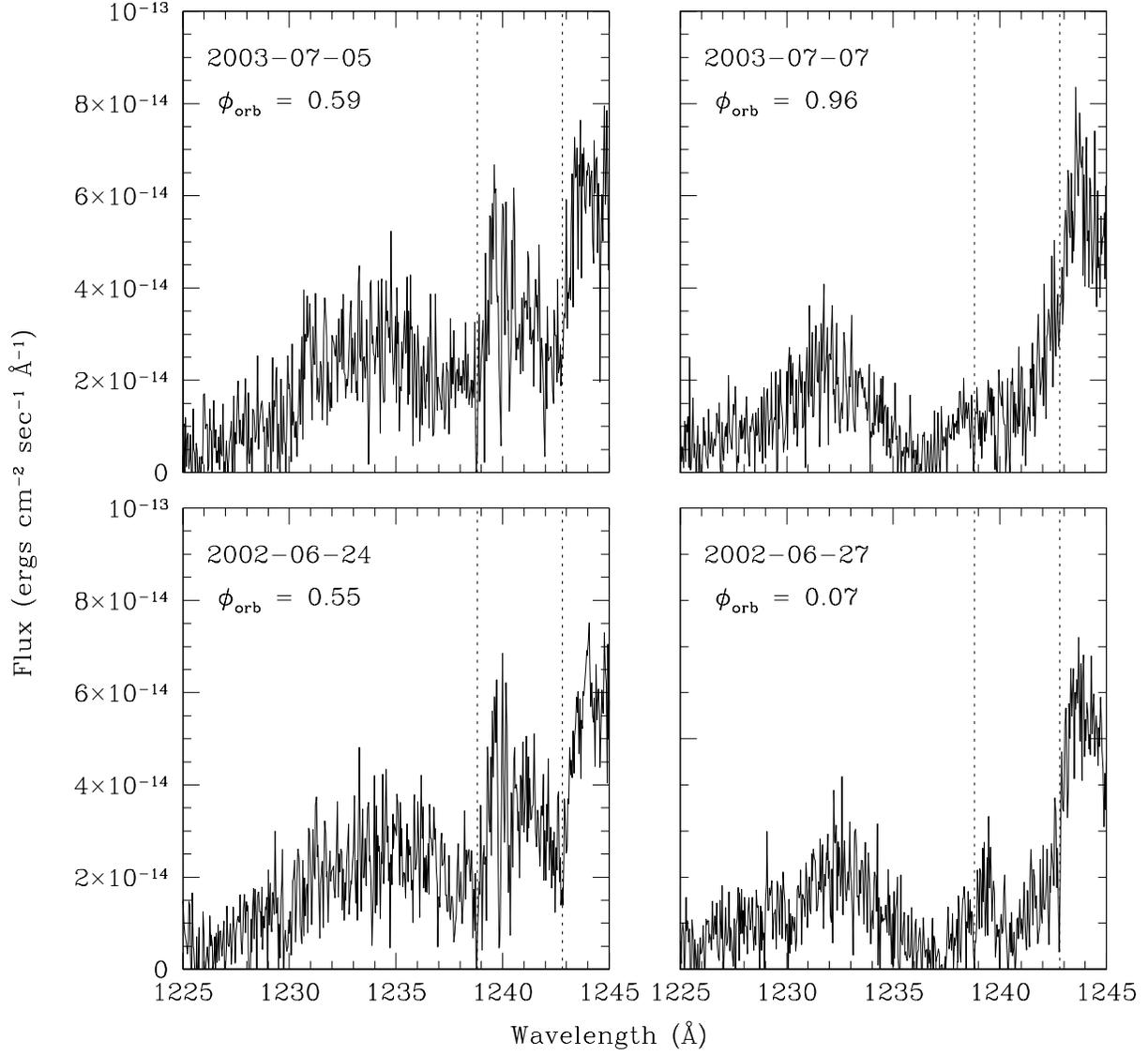}
\caption{Spectra of the N~V line at two different orbital phases and two diferent 
epochs. The dashed lines indicate the laboratory wavelengths for transitions 
listed in Table 2.} 
\label{cygx1nv} 
\end{figure}

\begin{figure}
\includegraphics[width=1.0\textwidth]{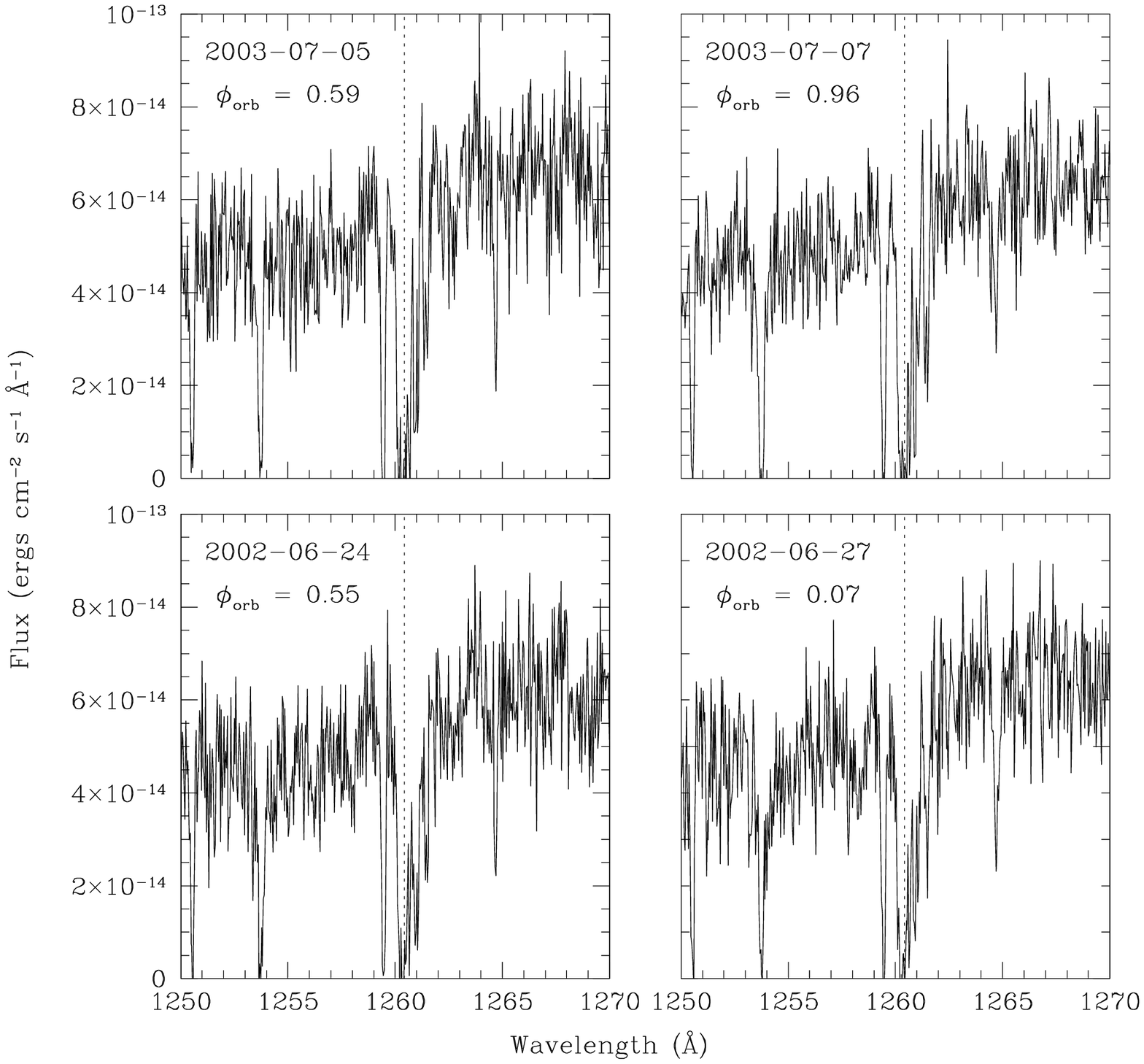}
\caption{As in figure \ref{cygx1nv} for Si~II}
\label{cygx1siii}
\end{figure}

\begin{figure}
\includegraphics[width=1.0\textwidth]{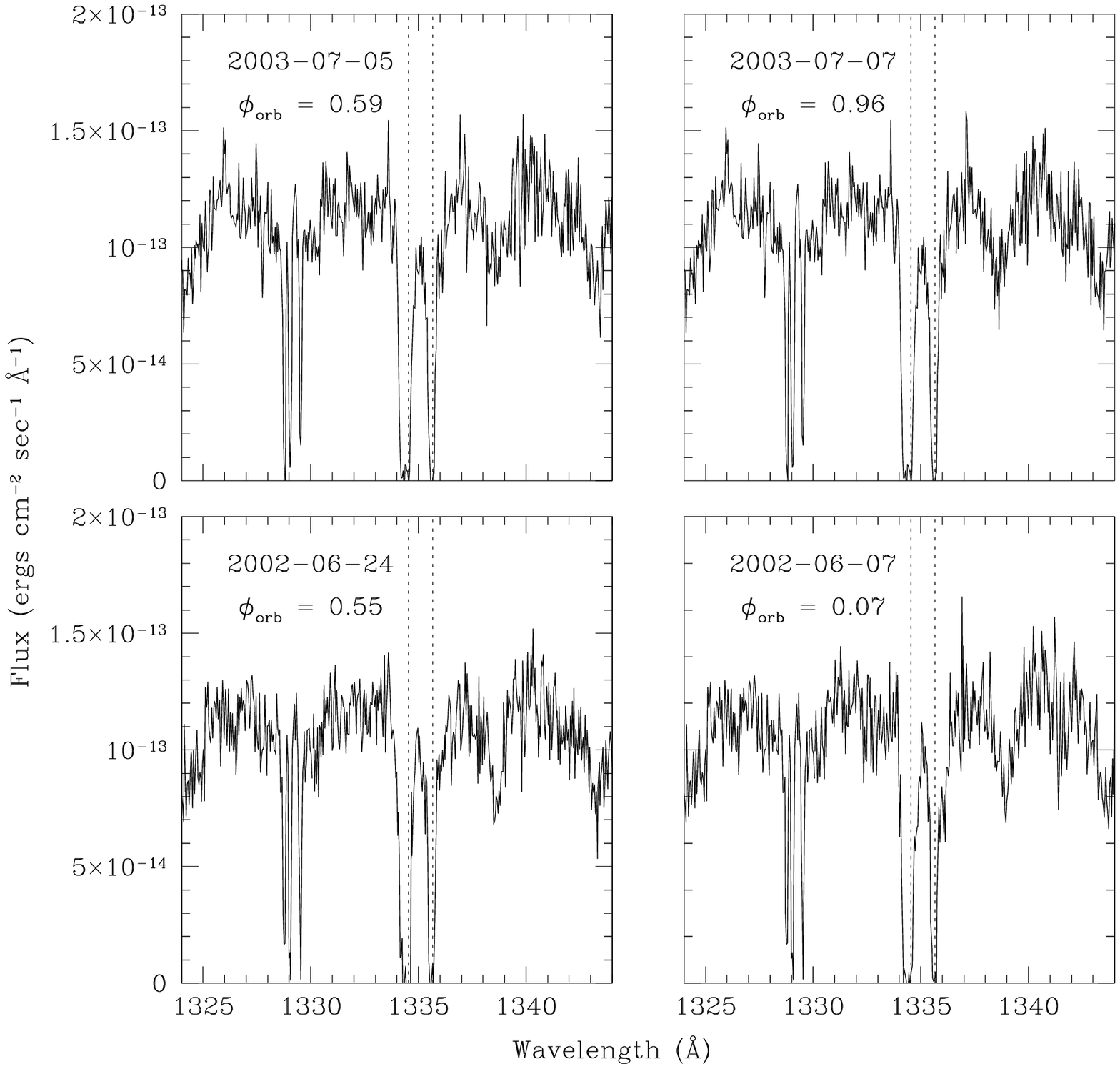}
\caption{As in figure \ref{cygx1nv} for C~II}
\label{cygx1cii}
\end{figure}

\clearpage

\begin{figure}
\includegraphics[width=1.0\textwidth]{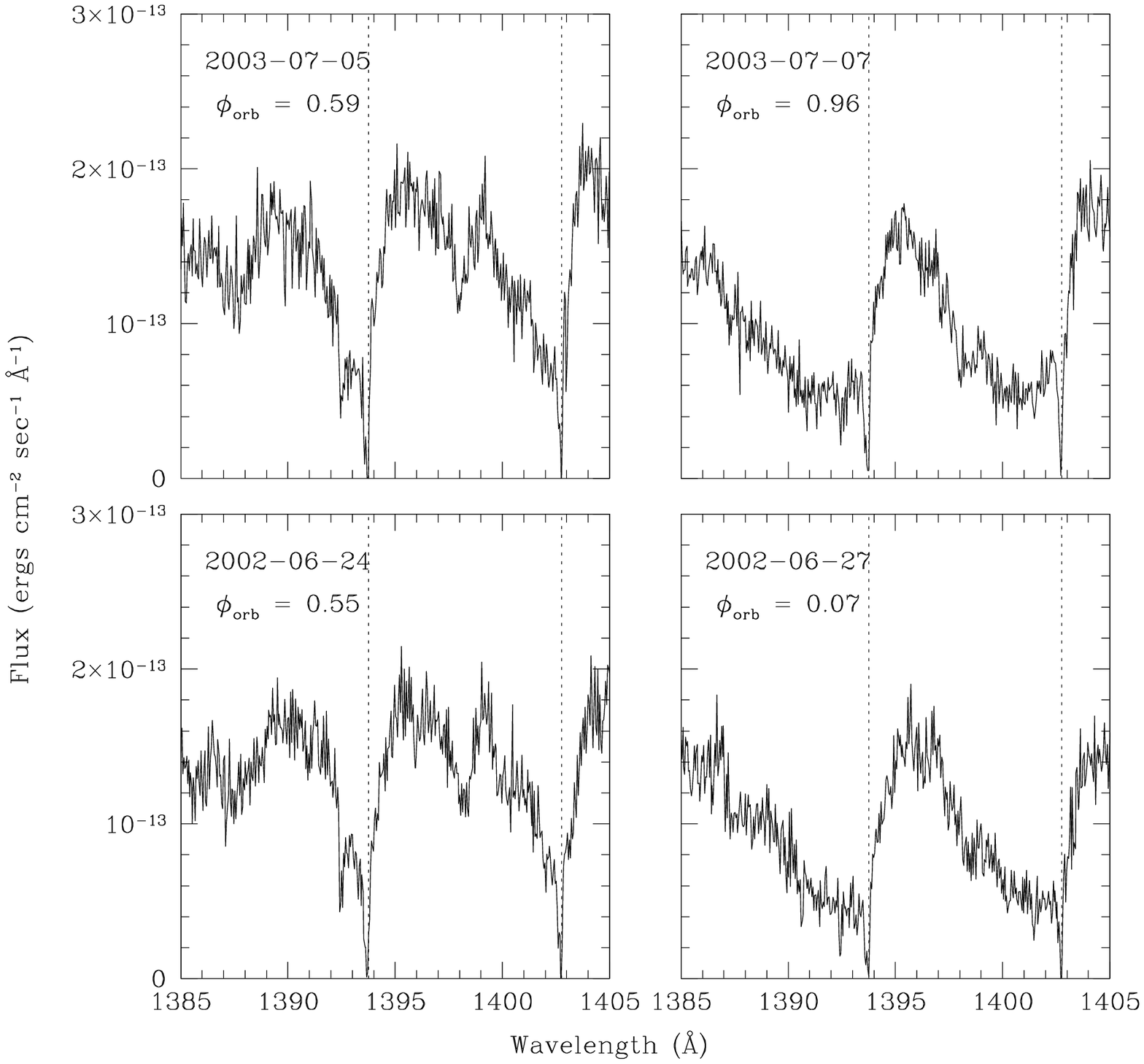}
\caption{As in figure \ref{cygx1nv} for Si~IV}
\label{cygx1siiv}
\end{figure}

\begin{figure}
\includegraphics[width=1.0\textwidth]{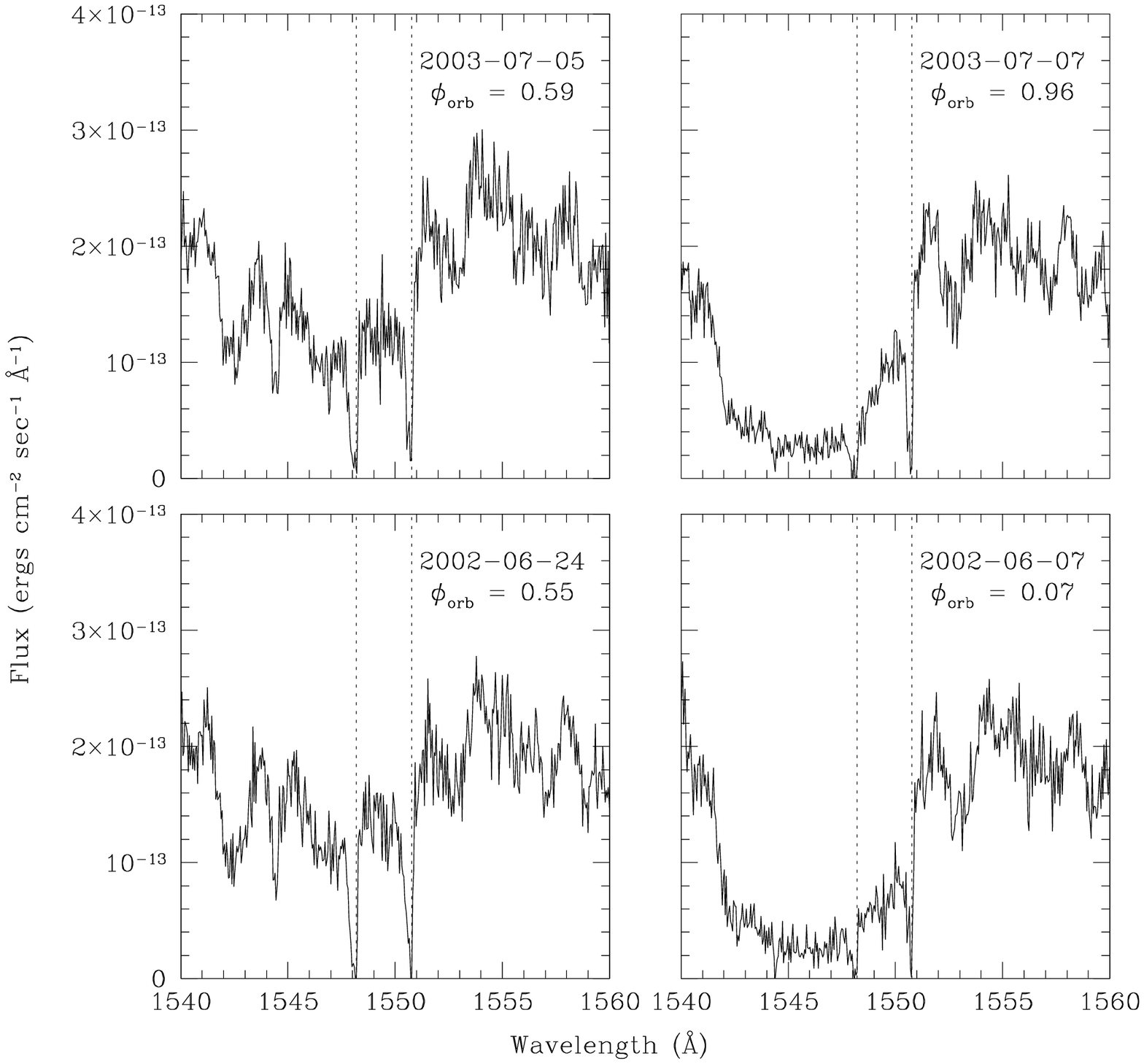}
\caption{As in figure \ref{cygx1nv} for C~IV}
\label{cygx1civ}
\end{figure}

\begin{figure}
\includegraphics[width=1.0\textwidth]{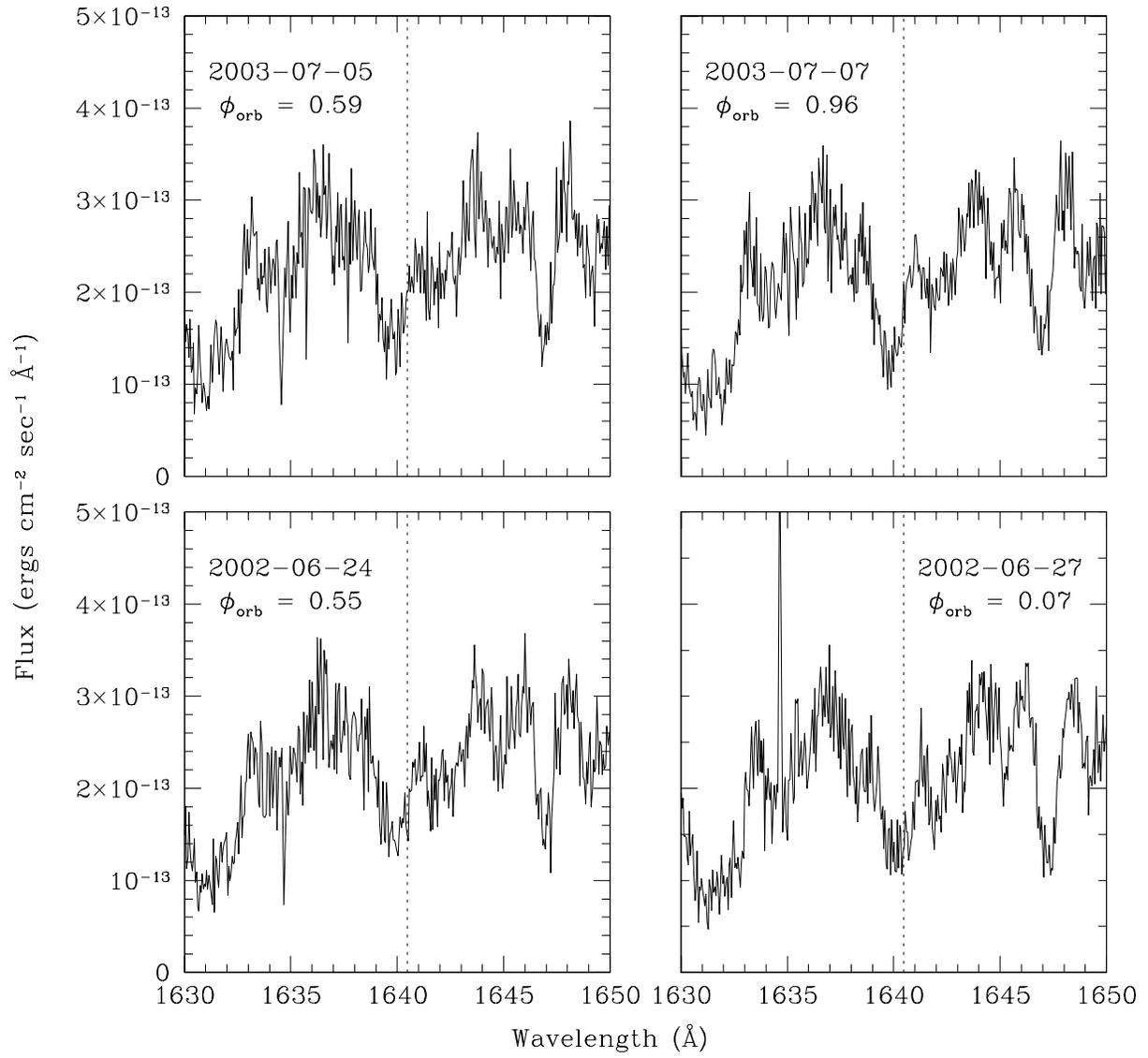}
\caption{As in figure \ref{cygx1nv} for HeII}
\label{cygx1heii}
\end{figure}

\clearpage

\begin{figure}
\includegraphics[width=1.0\textwidth]{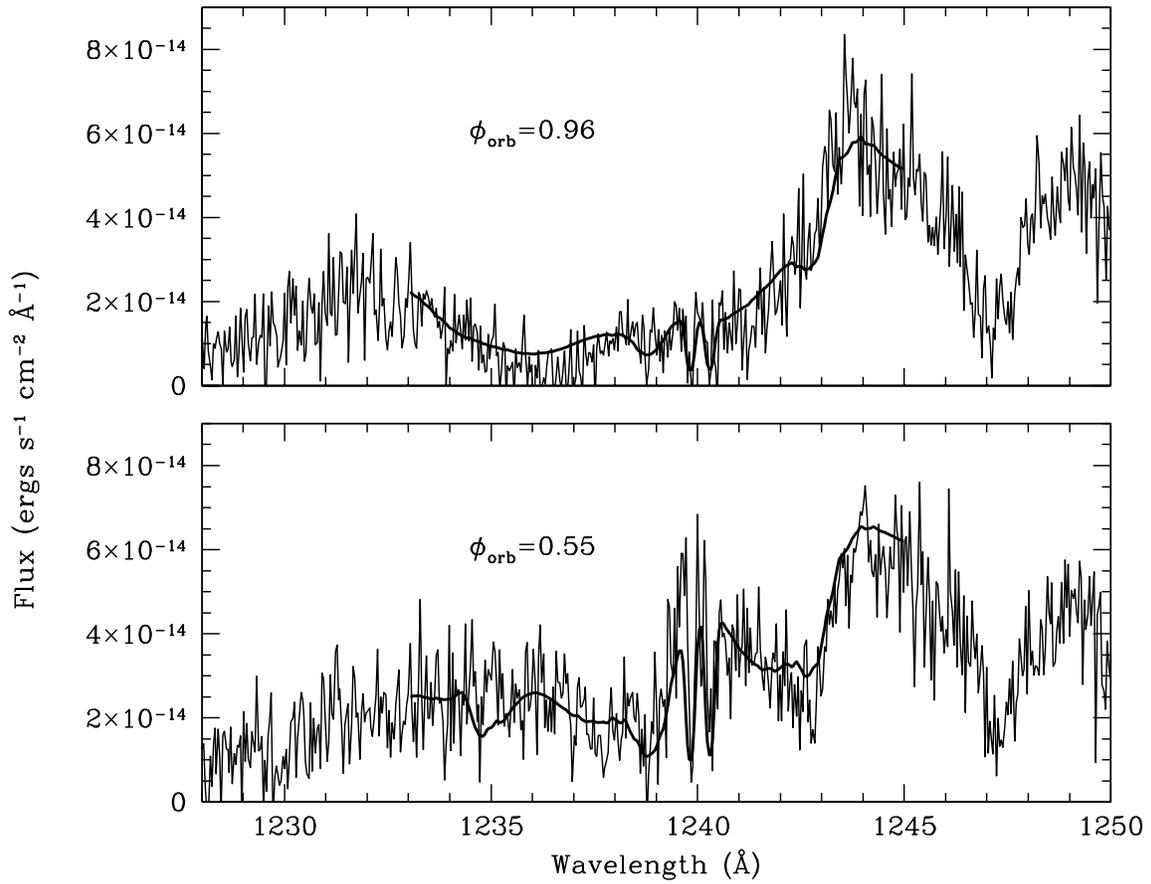}
\caption{Closeup of Cyg X-1 N~V line profiles.
The blue line depicts a model computed using the SEI 
method (Lamers, Cerruti-Sola, \& Perinotto 1987),
using a wind
velocity law given by $v_\infty~(1-1/r)^{\beta}$.} 
\label{nvmod}
\end{figure}

\begin{figure}
\includegraphics[width=1.0\textwidth]{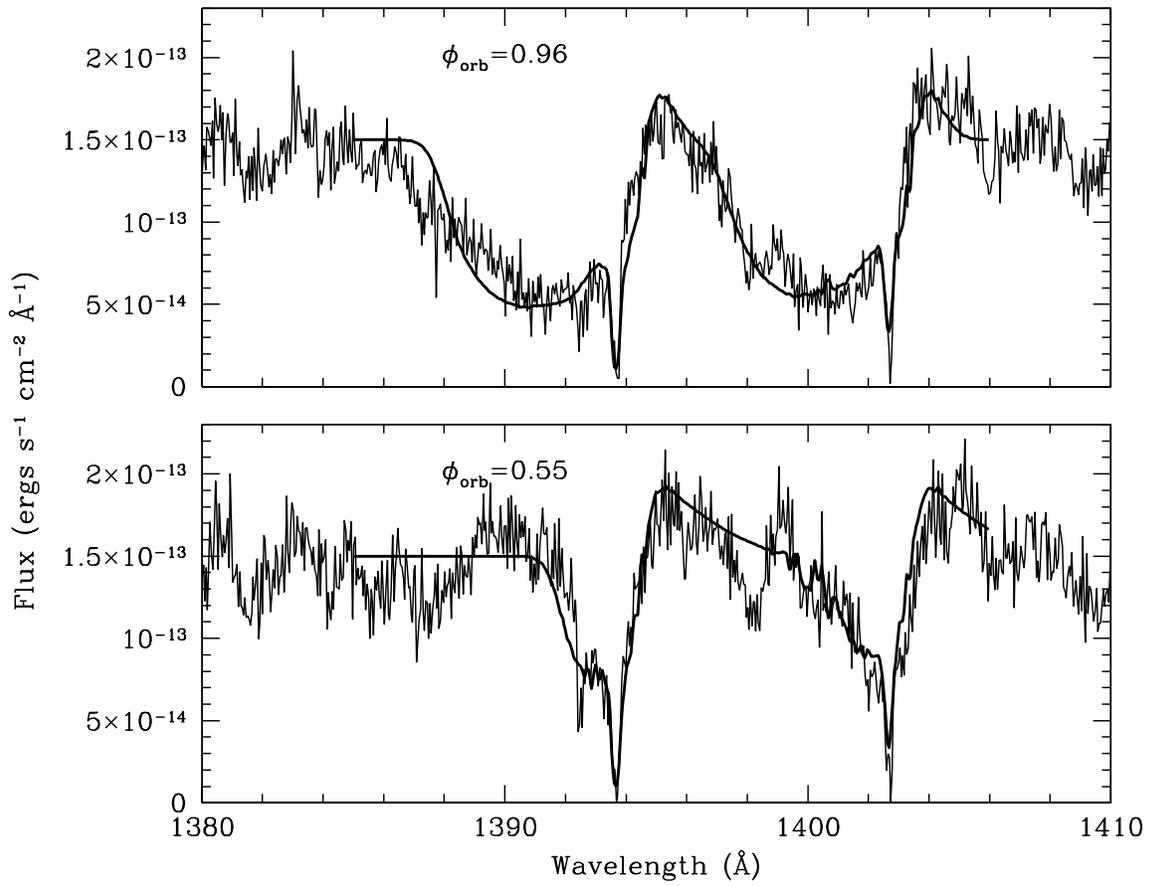}
\caption{As in Fig. \ref{nvmod} for Si~IV.} 
\label{siivmod}
\end{figure}

\clearpage

\begin{figure}
\includegraphics[width=1.0\textwidth]{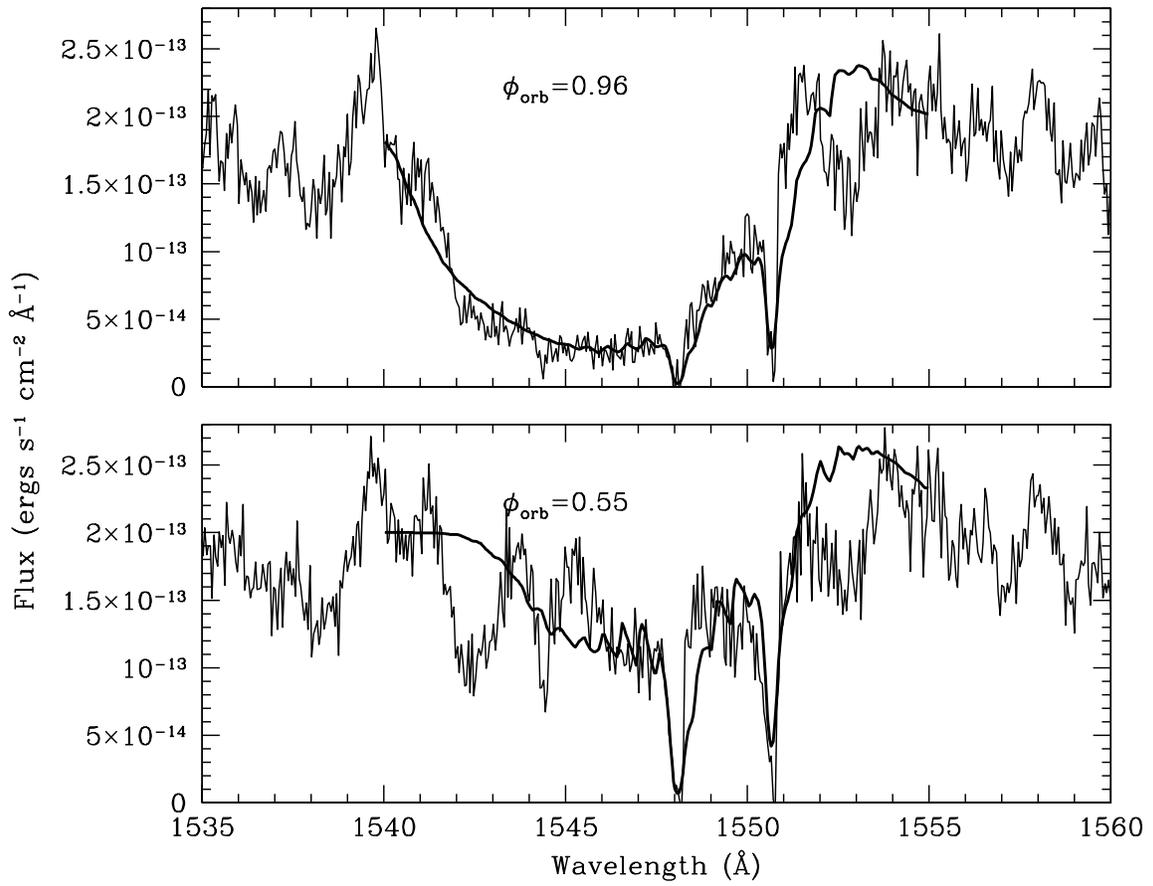}
\caption{As in Fig. \ref{nvmod} for C~IV.} 
\label{civmod}
\end{figure}

\begin{figure}
\includegraphics[width=1.0\textwidth]{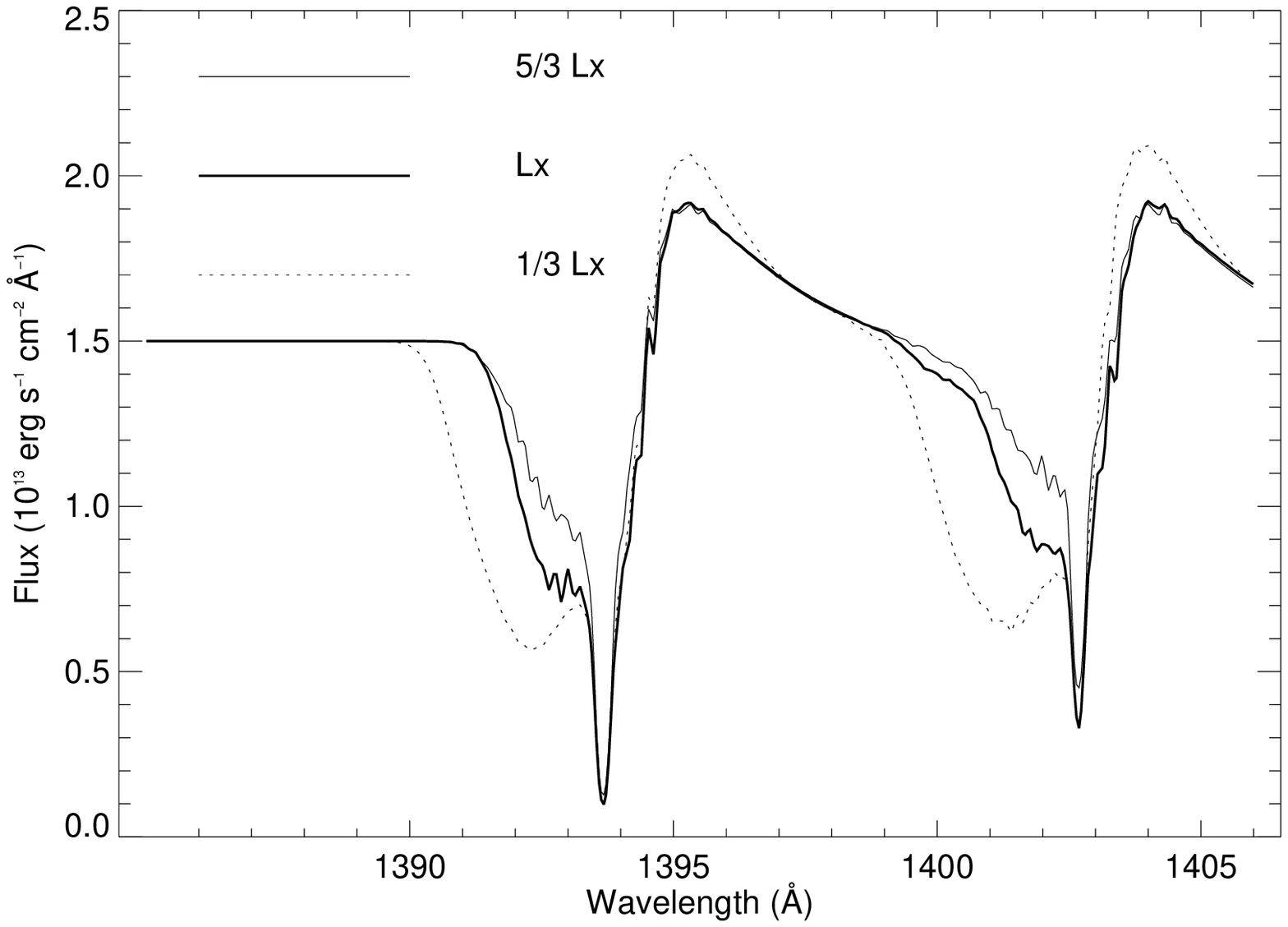}
\caption{The response of the wind-formed P~Cygni lines to X-ray
luminosities of $1/3 L_x$, $L_x$, and $5/3 L_x$
(dotted, bold, and regular thickness, respectively).
All other parameters are as assumed for the best fit to the Si~IV lines.
This figure shows the model at $\phi$=0.55 and ignores the 
effects of light travel-time, which
would tend to decrease the change in the line with
changing X-ray luminosity.}
\label{changlum1}
\end{figure}

\begin{figure} 
\includegraphics[width=1.0\textwidth]{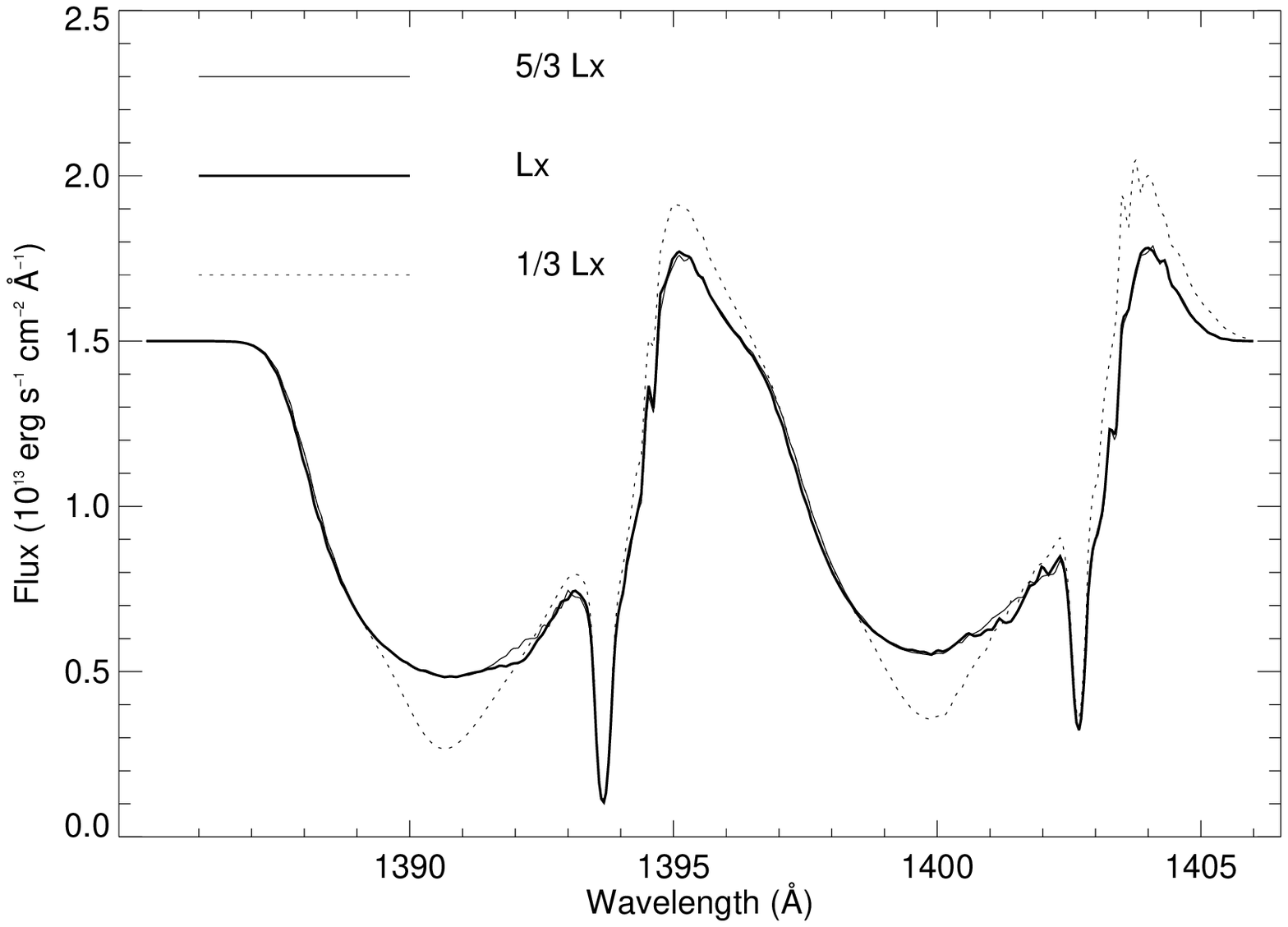}
\caption{The response of the wind-formed P~Cygni lines to X-ray
luminosities of $1/3 L_x$, $L_x$, and $5/3 L_x$
(dotted, bold, and regular thickness, respectively). All other
parameters are as assumed for the best fit to the Si~IV lines. This
figure shows the model at $\phi$=0.96 and ignores the effects of light
travel-time, which would tend to decrease the change in the line with
changing X-ray luminosity.}
\label{changlum2}
\end{figure}

\clearpage

\begin{figure} 
\includegraphics[width=1.0\textwidth]{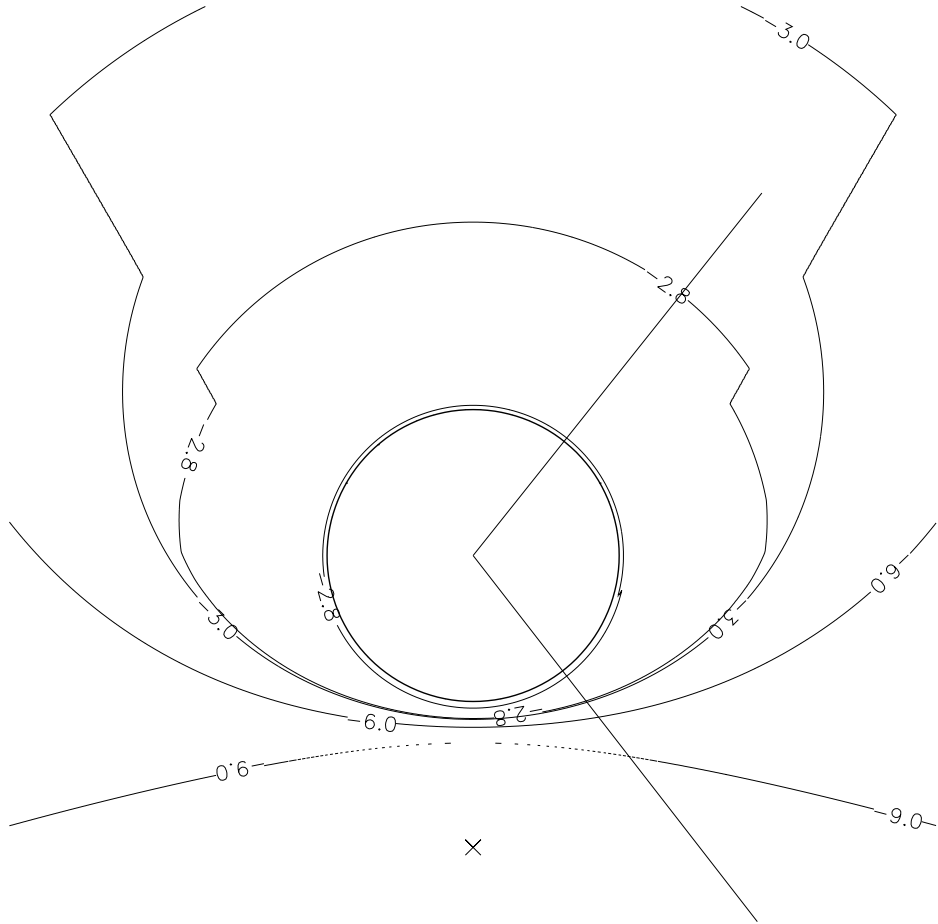}
\vspace{-3in}
\caption{Contours of Log$_{10}$ of Si~{\sc IV} abundance outside of
the shadow region. We assume the X-ray luminosity is 
$(1/3)1.6\times10^{37}$~erg~s$^{-1}$, and that all other parameters have their
best-fit values.  We include only ionization from the black hole and not
the ambient ionization of the wind.
The circle represents the supergiant
and the X the position of the compact source.}
\label{conplot1}
\end{figure}

\begin{figure} 
\includegraphics[width=1.0\textwidth]{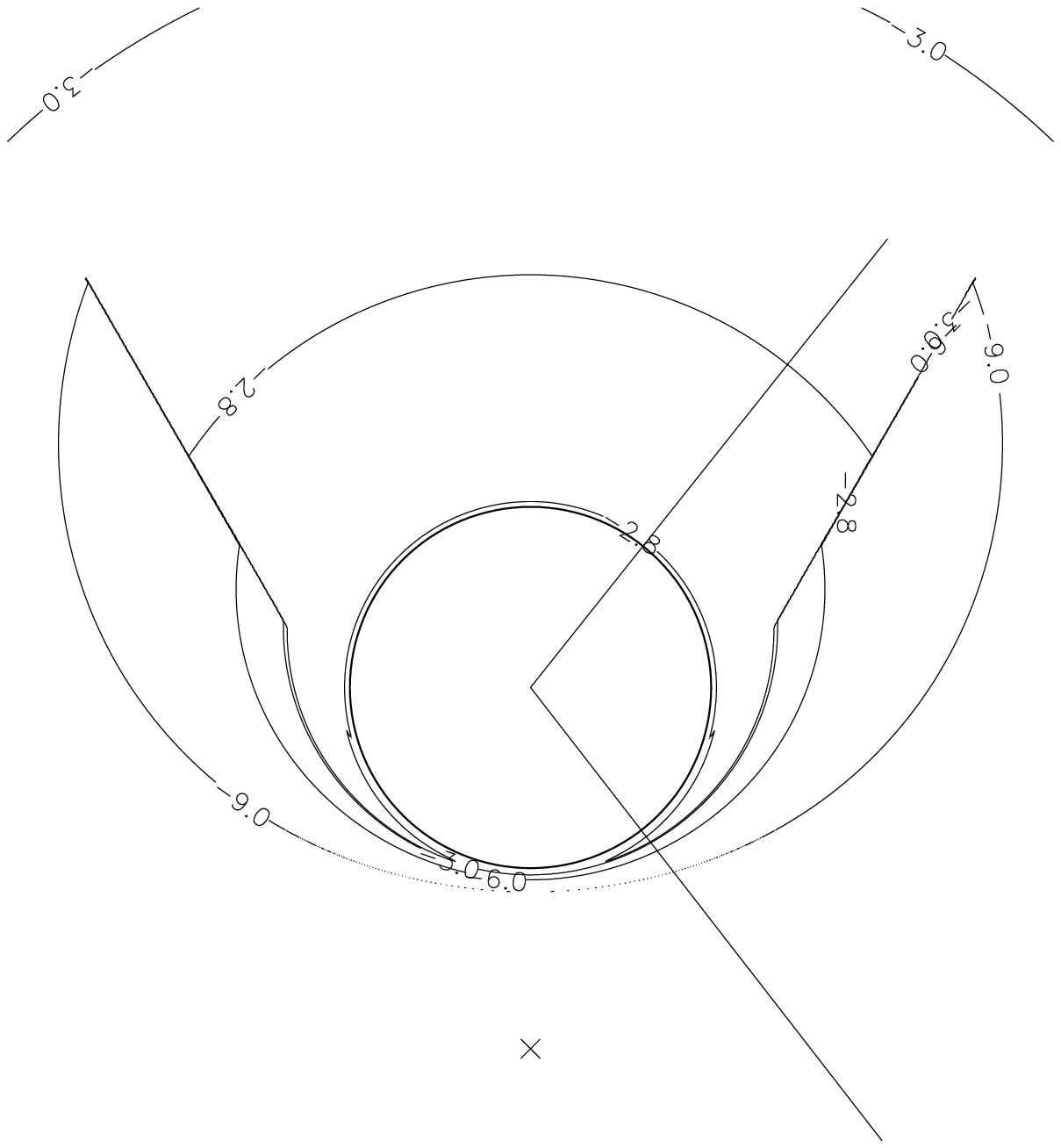}
\vspace{-3in}
\caption{Contours of Log$_{10}$ of Si\,{\sc IV} abundance outside of
the shadow region. We assume the X-ray luminosity is
$1.6\times10^{37}$~erg~s$^{-1}$, and that all other parameters have
their best-fit values.  We include only ionization from the black hole
and not the ambient ionization of the wind.}
\label{conplot2}
\end{figure}

\begin{figure} 
\includegraphics[width=1.0\textwidth]{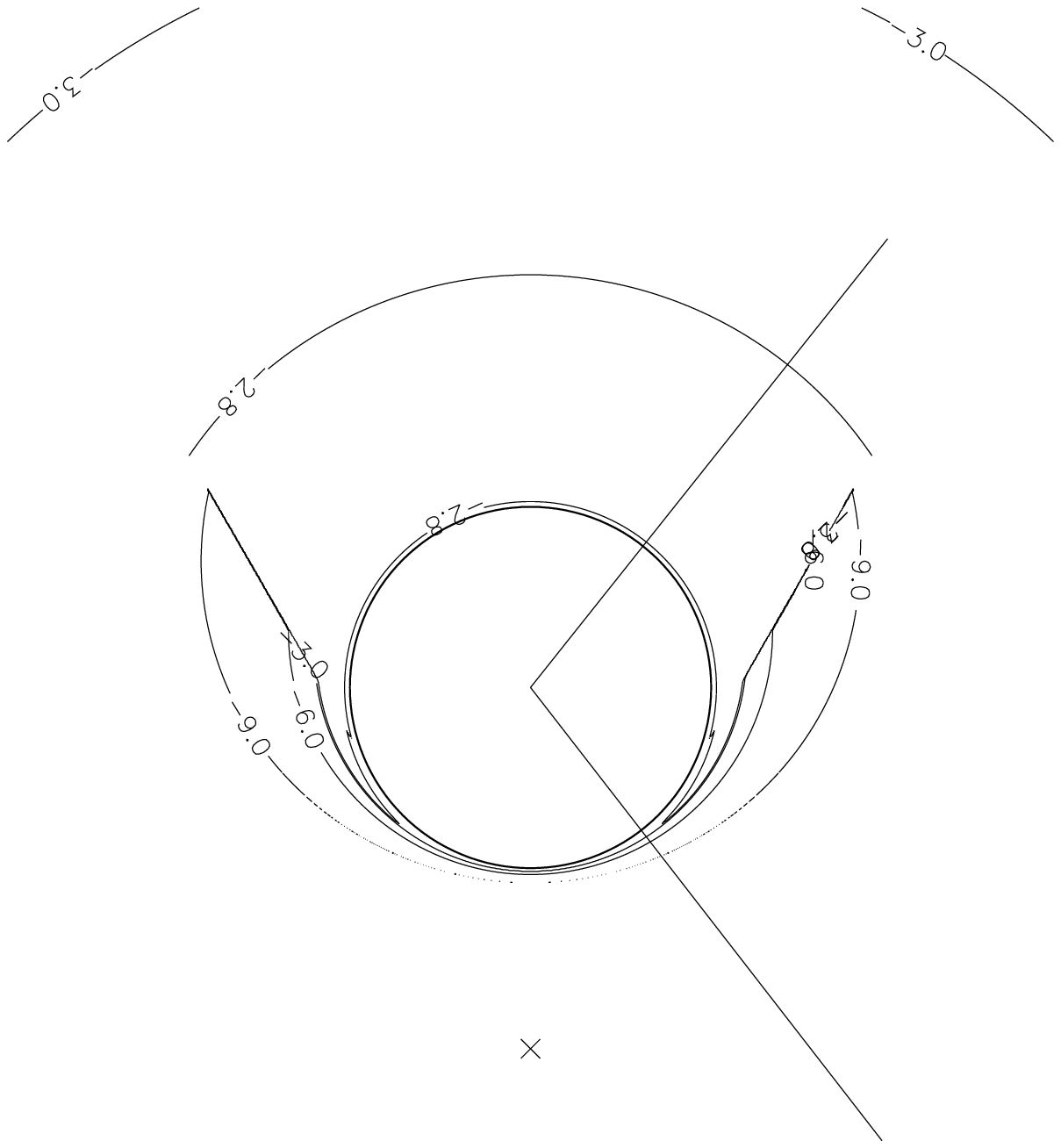}
\vspace{-3in}
\caption{Contours of Log$_{10}$ of Si\,{\sc IV} abundance outside of
the shadow region. We assume the X-ray luminosity is 
$(5/3)1.6\times10^{37}$~erg~s$^{-1}$, and that all other parameters have their
best-fit values.  We include only ionization from the black hole and not
the ambient ionization of the wind.}
\label{conplot3}
\end{figure}

\clearpage

\begin{figure} 
\includegraphics[width=1.0\textwidth]{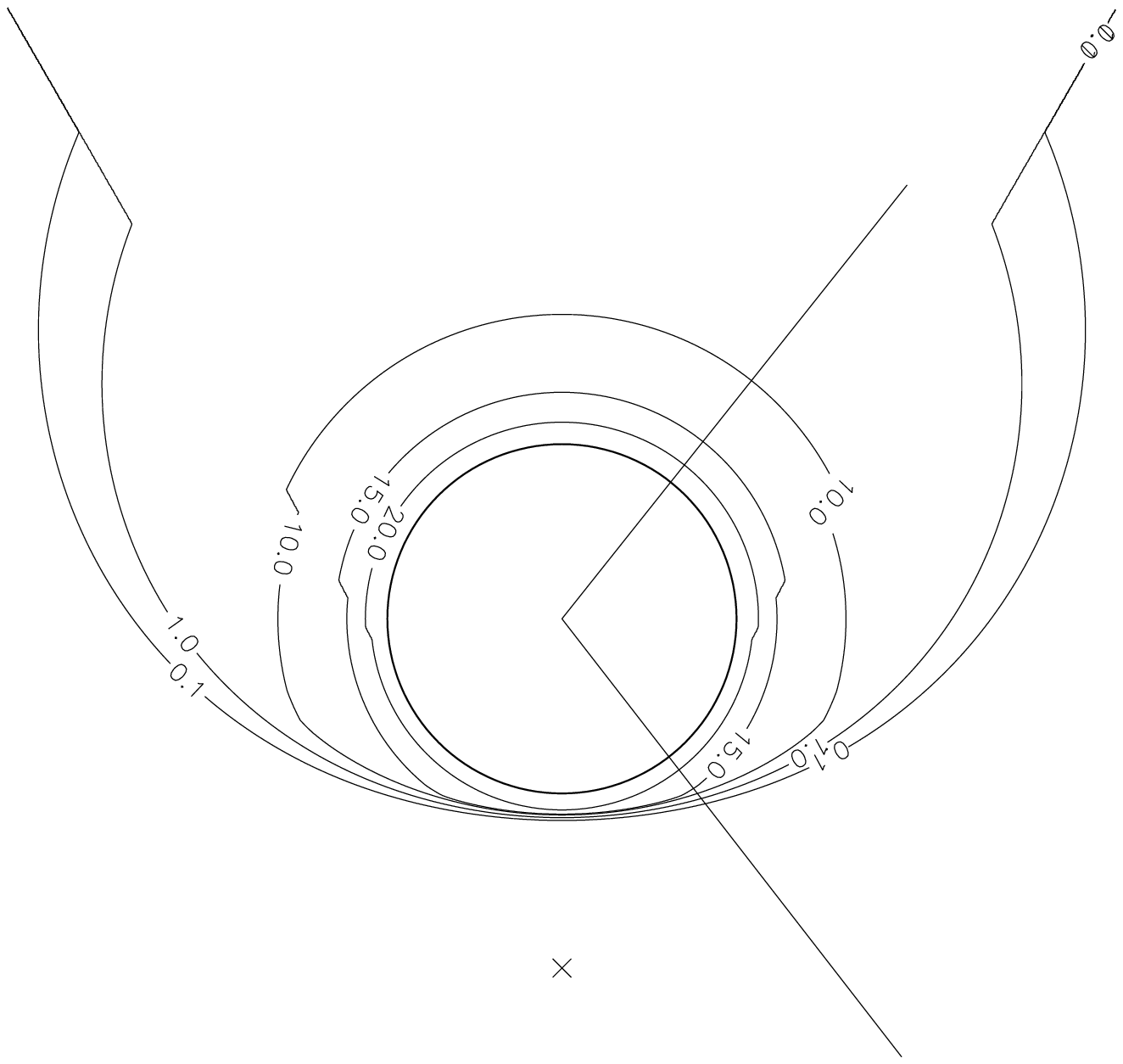}
\vspace{-3in}
\caption{Contours of radial optical depth $\tau$, given
the X-ray luminosity is 
$(1/3)1.6\times10^{37}$~erg~s$^{-1}$.}
\label{contau1}
\end{figure}

\begin{figure} 
\includegraphics[width=1.0\textwidth]{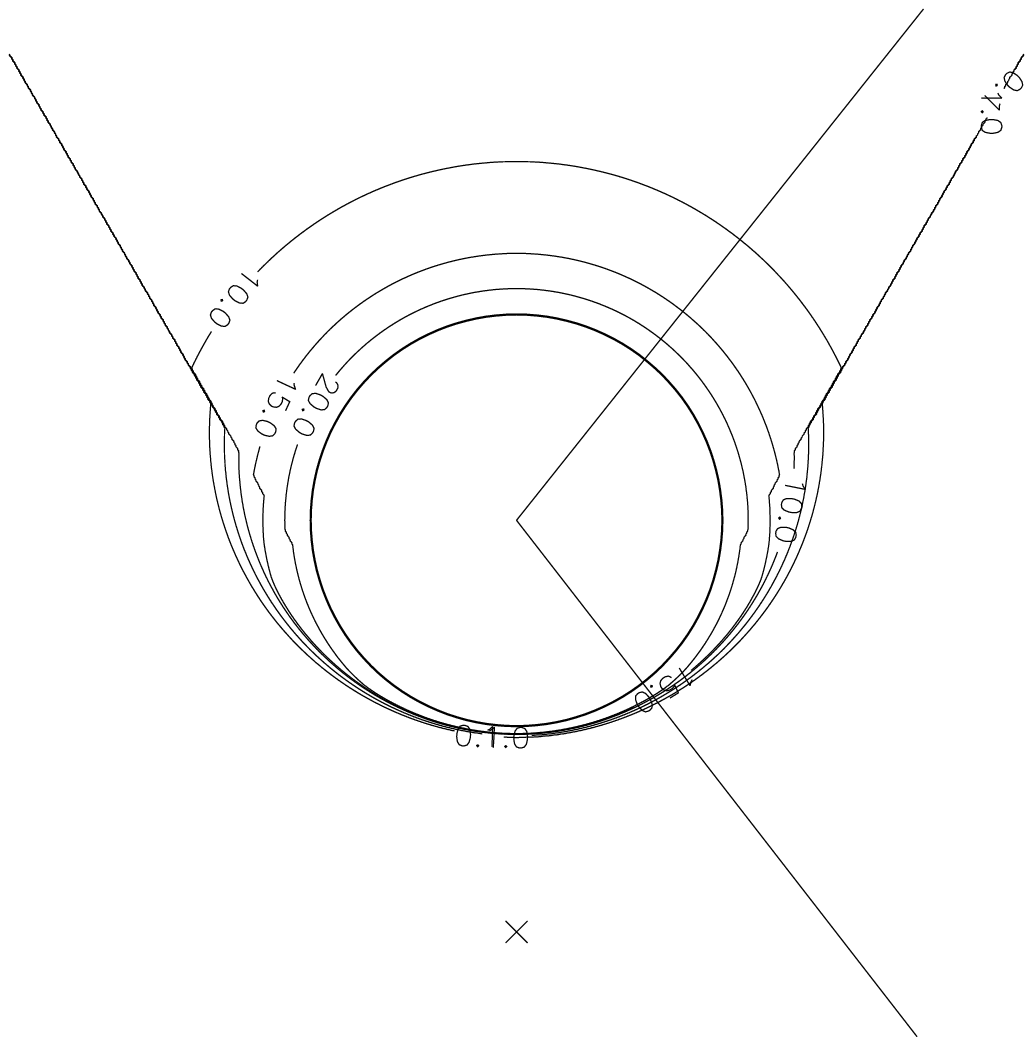}
\vspace{-3in}
\caption{Contours of radial optical depth $\tau$, given
the X-ray luminosity is 
$1.6\times10^{37}$~erg~s$^{-1}$.}
\label{contau2}
\end{figure}

\begin{figure} 
\includegraphics[width=1.0\textwidth]{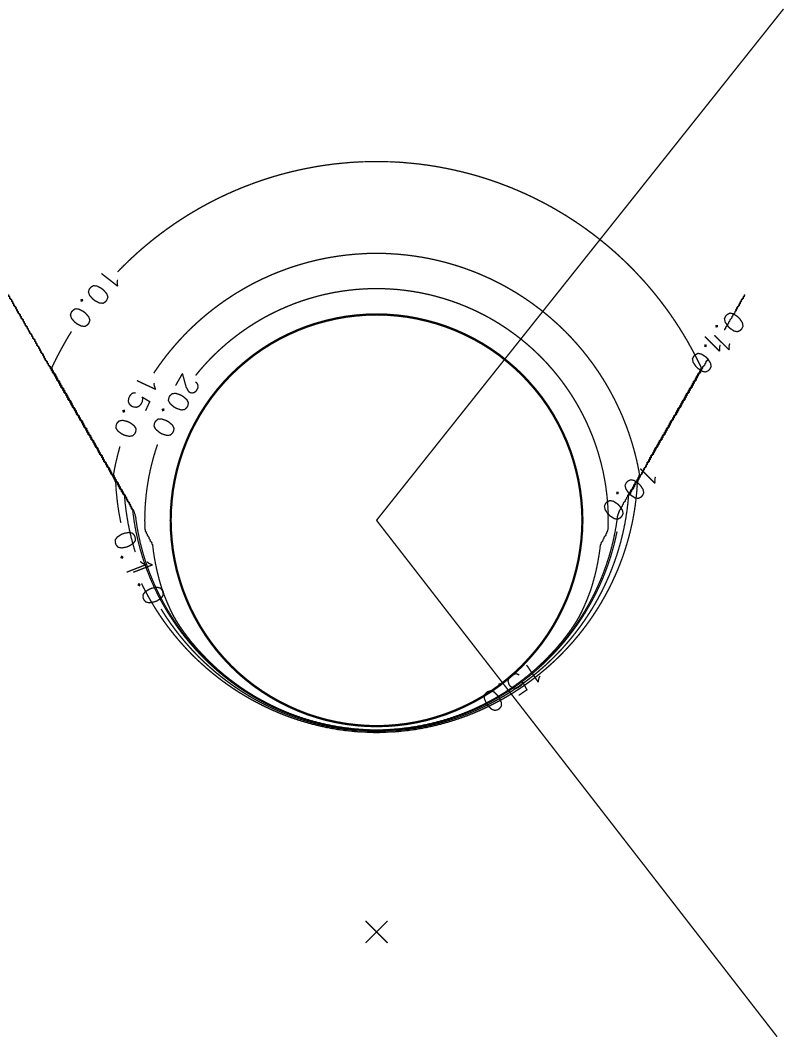}
\vspace{-3in}
\caption{Contours of radial optical depth $\tau$, given
the X-ray luminosity is 
$(1/3)1.6\times10^{37}$~erg~s$^{-1}$.}
\label{contau3}
\end{figure}

\end{document}